\documentclass[a4paper,11pt]{article}
\usepackage[dvips]{graphicx}
\usepackage{amsmath}        
\usepackage{amssymb}        
\usepackage{cite}
\usepackage{cancel}

\begin{document}

\title{A Supersymmetric Grand Unified Model with Noncompact Horizontal Symmetry}

\author{Naoki Yamatsu\footnote{Electronic address: nyamatsu@indiana.edu}\\
{\it\small Department of Physics, Indiana University, Bloomington, IN 47405, USA}}
\date{ }

\maketitle

\begin{abstract}
 In a supersymmetric $SU(5)$ grand unified model with a  horizontal
 symmetry $SU(1,1)$, we discuss spontaneous generation of generations
 to produce three chiral generations of quarks and leptons and one
 generation of higgses by using one structure field with a half-integer
 spin of $SU(1,1)$ and two structure fields with integer spins.
 In particular, the colored higgses can disappear without
 fine-tuning. The difference of the Yukawa coupling  matrices between
 the down-type quarks and charged leptons is discussed.
 We show that some special $SU(1,1)$ weight assignments include
 $R$-parity as a discrete  subgroup, and $R$-parity remains even after
 we take into account the $SU(1,1)$  breaking effects from all the VEVs
 of the structure and matter fields. The assignments forbid the
 baryon and/or lepton number violating terms except a superpotential
 quartic term including a coupling of two lepton doublets and
 two up-type  higgses. We discuss how to generate sizable neutrino masses.
 We show that the proton decay derived from the colored higgses is
 highly suppressed.  
\end{abstract}

\section{Introduction}
 
The implications for the fundamental theory of nature from low energy
phenomena come from the problems of 't Hooft's naturalness
\cite{'tHooft:1980xb} 
and fine-tuning \cite{Dimopoulos:1981zb,Barbieri:1987fn}.
The problems are the window to hidden structures of nature.  One of the
naturalness problems of hierarchical mass structures of quarks and
leptons suggests the existence of horizontal symmetry
\cite{Wilczek:1978xi,Froggatt:1978nt,Yanagida:1979as,Maehara:1979kf}.
The strong $CP$ problem 
\cite{Belavin:1975fg,'tHooft:1976up,Jackiw:1976pf,Callan:1976je} 
implies the spontaneous discrete symmetry breaking containing $P$/$CP$
symmetry 
\cite{Nelson:1983zb,Nelson:1984hg,Barr:1984fh,Mohapatra:1978fy,Kuchimanchi:1995rp,Mohapatra1996,Babu2002,Kuchimanchi:2010xs,Kuchimanchi:2012xb}
or existence of the axion
\cite{Peccei:1977hh,Peccei:1977ur,Weinberg:1977ma,Wilczek:1977pj,Kim:1979if}.
For a review, see, e.g., Ref.~\cite{Cheng:1987gp}.
The fine-tuning problem of quadratic divergence of the higgs mass term
suggests the existence of 
supersymmetry (SUSY) \cite{Wess:1971yu,Wess:1992cp,Nilles1984},
extra-dimension \cite{Hatanaka:1998yp,ArkaniHamed:1998rs,Randall:1999ee},
or technicolor \cite{Weinberg:1979bn,Susskind:1978ms}.

The implications are not only from the issues of naturalness, but also the
quantization of charges, the anomaly cancellation of the standard model
(SM) gauge groups $SU(3)_C\times SU(2)_L\times U(1)_Y (=G_{\rm SM})$ by
each generation of quarks and leptons at low energies
\cite{Bouchiat:1972iq}, 
the unification of three gauge coupling constants at the unification
scale, and the matter unification of quarks and leptons in SM for one or
two representations in grand unified groups. 
They seem to suggest that one of the hidden structures of nature 
is some unified gauge symmetry 
\cite{Pati:1974yy,Georgi:1974sy}.
As is well-known, candidates for the grand unified gauge symmetry are 
simple groups, such as 
$SU(5)$ \cite{Georgi:1974sy,Koh:1982zz,DelOlmo:1987bc}, 
$SU(6)$ \cite{Inoue:1977qd,Hartanto2005}, 
$SO(10)$ \cite{Fritzsch:1974nn,Koh:1984vs,Nath2001,Fukuyama2005}, 
and $E_6$ 
\cite{Gursey:1975ki,Anderson:1999em,Anderson:2000ni,Anderson:2001sd}.
For a review, see e.g.,
Refs.~\cite{GellMann:1976pg,Slansky:1981yr,Georgi:1982jb}. 
Any grand unified model explains the quantization of charges, and some
of them explain the anomaly cancellation and the SM gauge coupling
unification. Here we will focus on the $SU(5)$ unified group.

As is well-known, the non-supersymmetric $SU(5)$ grand unified model
\cite{Georgi:1974sy}
that contains the minimal numbers of quarks, leptons, and higgs predicts
rapid proton decay via $X$ and $Y$ gauge bosons.
As long as the colored higgs mass is $O(M_{\rm GUT})$, since the Yukawa
coupling 
constants of the first and second generations of quarks and leptons
coupling to the colored higgs are smaller than the gauge coupling
constant,
the most strict restriction for the proton decay via
the $X$ and $Y$ gauge bosons comes from the mode $p\to \pi^0e^+$. 
By using the chiral Lagrangian technique, the lifetime is given by
\cite{Hisano:1992jj}
\begin{align}
\tau(p\to\pi^0e^+)\to 1.1\times 10^{36}\times
\left(\frac{M_V}{10^{16}\mbox{GeV}}\right)^4
\left(\frac{0.003\mbox{GeV}^3}{\alpha}\right)^2\ \mbox{years},
\label{proton-decay-XY}
\end{align}
where $M_V$ is the $X$ and $Y$ gauge boson mass and $\alpha$ is
a hadron matrix element.
When we use the gauge bosons masses 
$M_V\sim M_{\rm GUT}\sim 10^{15}$ GeV and 
a hadron matrix element $\alpha=0.003$ GeV$^3$,
we obtain $\tau(p\to \pi^0e^+)=1.1\times 10^{32}$ years.
From the latest result from the super-Kamiokande \cite{Nishino:2012ipa}, 
the lifetime $\tau(p\to\pi^0e^+)>8.2\times 10^{33}$ years at 90 \%C.L. 
Thus, as is well-known, the non SUSY $SU(5)$ GUT model seems to be ruled
out.

Fortunately, in the minimal $SU(5)$ SUSY GUT model 
\cite{Ellis:1981tv,Sakai:1981pk,Weinberg:1981wj,Dimopoulos:1981zb}
the GUT scale $M_{\rm GUT}$ becomes $O(10^{16})$ GeV. 
Substituting $M_V= 10^{16}$ GeV in Eq.~(\ref{proton-decay-XY}), 
we obtain the proton lifetime $\tau(p\to \pi^0e^+)=1.1\times 10^{36}$
years. Thus, the lifetime satisfies the current bound.
However, it is also known that the minimal $SU(5)$ SUSY GUT model
suffers from rapid proton decay induced from the colored higgses 
\cite{Nath:1985ub,Goto:1998qg,Hisano:1992jj,Hisano:2000dg}.
According to Ref.~\cite{Goto:1998qg}, the colored higgs masses must
be greater than $10^{17}$ GeV for any $\tan\beta$ by using 
the recent super-Kamiokande result for the lifetime 
$\tau(p\to K^+\bar{\nu})>3.3\times 10^{33}$ years at 90\% C.L.
\cite{Miura:2010zz} when we assume that the sfermion masses are less
than 1 TeV.
Thus, the colored higgses must have the effective mass greater than 
$O(10^{17})$ GeV. 
On the other hand, the doublet higgs must have
$O(m_{\rm SUSY})$. This is known as a doublet-triplet splitting problem
\cite{Masiero:1982fe,Grinstein:1982um,Inoue:1985cw,Witten:1985xc,Yanagida:1994vq,Kawamura:2000ev,Maekawa:2002bk,Bando2003}.

In addition, the minimal $SU(5)$ GUT model gives an unacceptable
relation between the Yukawa coupling constants of down-type quarks and
charged leptons without taking into account the higher dimensional
operators including the nonvanishing VEVs of the adjoint representation.
To break the minimal GUT relation of the Yukawa coupling constants
between down-type quarks and charged leptons, roughly speaking, we can
classify two methods; one is to consider the higher dimensional operators
including the $SU(5)$ adjoint higgs field; another is to introduce the
higher dimensional representations, such as Georgi-Jarlskog manner
(see, e.g., Ref.~\cite{Georgi:1979df} for an $SU(5)$ Non-SUSY GUT model
and Ref.~\cite{Murayama1992} for an $SU(5)$ SUSY GUT model).
The above ways can also be mixed.

Even when we consider SUSY GUT models, they do not give us any insight
about the hierarchy of the Yukawa couplings and the number of chiral
generations of quarks, leptons and higgses.
The mass parameters at the GUT scale in the minimal supersymmetric
standard model (MSSM) \cite{Martin:1997ns,Chung:2003fi,Baer2006}
are given by Ref.~\cite{Ross:2007az} for several values of $\tan\beta$
by using the renormalization group equations of the two-loop gauge
couplings and the two-loop Yukawa couplings assuming an effective SUSY
scale of 500 GeV. For $\tan\beta=(10,38,50)$, the coupling constants of
the third generation of the up-type quark, the down-type quark, and the
charged lepton at the GUT scale are $y_t\simeq(0.48,0.49,0.51)$,
$y_b\simeq(0.051,0.23,0.37)$, and $y_\tau\simeq(0.070,0.32,0.51)$,
respectively. 
When we normalize the Yukawa coupling constants of the third
generations equal to one, the mass parameters of first, second and third
generations of the up-type quark, the down-type quark, and the charged
lepton for $\tan\beta=10$ are $(\tilde{y}_1,\tilde{y}_2,\tilde{y}_3)
\simeq(6.7\times 10^{-6},2.5\times 10^{-3},1)$, 
$(1.0\times 10^{-3},1,2\times 10^{-2},1)$, and  
$(2.5\times 10^{-4},6\times 10^{-2},1)$, respectively, where 
the subscript of $\tilde{y}_a$ $(a=1,2,3)$ stands for the generation
number. The values are almost the same for $\tan\beta=38$ and $50$.

The hierarchical structures of the Yukawa couplings of quarks and
leptons strongly suggest the existence of a hidden structure of nature. 
There have been many attempts to understand the origin of
the hierarchical structures and/or generations by using horizontal
symmetries $G_H$ 
\cite{Wilczek:1978xi,Froggatt:1978nt,Yanagida:1979as,Maehara:1979kf}:
e.g., non-abelian group symmetries
\cite{Ishimori:2010au,King:2001uz,Krnjaic:2012aj}, 
an abelian group $U(1)$
\cite{Froggatt:1978nt,Maekawa2004}, and
a noncompact nonabelian group symmetry $SU(1,1)$
\cite{Inoue:1994qz},
where the noncompact group $SU(1,1)$ is a special pseudo-unitary group
\cite{Gourdin:1967,Gilmore:102082}. 

In this article, we discuss an $\mathcal{N}=1$ supersymmetric vectorlike
$SU(5)$ GUT model with a noncompact horizontal symmetry $SU(1,1)$ to
solve the above problems. 
We summarize the main results of previous studies of $\mathcal{N}=1$
supersymmetric vectorlike models with a horizontal symmetry $SU(1,1)$
\cite{Inoue:1994qz,Inoue:2000ia,Inoue:2003qi,Yamatsu:2007,Yamatsu:2008,Yamatsu:2012}.
The number of chiral generations of matter fields, such as quarks,
leptons and higgses are determined by the spontaneous symmetry breaking
of the horizontal symmetry $SU(1,1)$, called the spontaneous generation
of generations \cite{Inoue:1994qz}.
Through the mechanism, the doublet-triplet splitting of higgses can be
realized without fine-tuning and also unreasonably suppressed tiny mass
parameters \cite{Inoue:2000ia,Yamatsu:2007}.
When the horizontal symmetry is unbroken, the original Yukawa coupling
matrices of matter fields are completely determined by $SU(1,1)$
symmetry. The Yukawa coupling constants of the chiral matter fields at low
energy are controlled by the $SU(1,1)$ symmetry and the $SU(1,1)$
breaking vacua. Each structure of Yukawa couplings of three chiral
generations of quarks and leptons has hierarchical structure
\cite{Inoue:1994qz,Inoue:2000ia,Yamatsu:2007,Yamatsu:2012}.
The problematic superpotential cubic terms
$\hat{Q}\hat{L}\hat{D}^c$, $\hat{D}^c\hat{D}^c\hat{U}^c$,
$\hat{L}\hat{L}\hat{E}^c$ are automatically forbidden,
where in the MSSM these terms are forbidden by $R$-parity
\cite{Fayet:1977yc} (or matter parity \cite{Bento:1987mu}) 
to prevent rapid proton decay (For a review, see, e.g.,
Ref.~\cite{Barbier:2004ez}). 
The dangerous superpotential quartic terms
$\hat{Q}\hat{Q}\hat{Q}\hat{L}$ and
$\hat{U}^c\hat{U}^c\hat{D}^c\hat{E}^c$ are also not allowed 
where the usual $R$-parity cannot forbid these terms
\cite{Inoue:2000ia,Yamatsu:2007}.

We now discuss $\mathcal{N}=1$ supersymmetric noncompact gauge theory
since our model is based on an $\mathcal{N}=1$ supersymmetric  noncompact
gauge theory. As is well-known, renormalizable noncompact gauge theories
have ghost problems; at least one gauge field has a negative metric in
the canonical kinetic term, which indicates the wrong sign and this is
physical ghost; the structure fields belonging to the finite
dimensional representations also have the physical ghosts.  
A solution of this problem, discussed in Ref.~\cite{Yamatsu:2009}, is
to use an $\mathcal{N}=1$ supersymmetric model with a noncompact gauge
group $SU(1,1)$ that has noncanonical K\"ahler function and gauge kinetic
function with linear representation of $SU(1,1)$ gauge transformation. 
At least at classical level, the Lagrangian has gauge and K\"ahler
metrics positive definite at proper vacua, and thus no ghost fields
exist at the vacua. For another solution of this problem, see, e.g.,
Refs.~\cite{Cremmer:1979up,deWit:1983xe}.

The main purpose of this paper is to show that an $SU(5)$ SUSY GUT
model with the noncompact horizontal symmetry $SU(1,1)$ naturally
satisfies current proton decay experiments, solves the doublet-triplet
mass splitting problem, and avoids the unrealistic GUT relation for
Yukawa couplings. 
In addition, we will see that this model can accommodate $R$-parity as
a discrete subgroup of the horizontal symmetry.

Here we clarify the difference between this work and the previous
works with models with the noncompact horizontal symmetry. This is the
first trial to construct a concrete $SU(5)$ model.
We apply the spontaneous generation of generations for the model with
the matter content of an $SU(5)$ grand unified model. 
The mixing structures of quarks and leptons 
that represent the ratio of the mixing between each chiral mode and the
components of matter fields are basically Type-I, II,
and III structures discussed in Ref.~\cite{Yamatsu:2012},
where structure fields are chiral
superfields with the finite dimensional representation of $SU(1,1)$.
Since the discussion in Ref.~\cite{Yamatsu:2012}
is the simplest case that contains
only two structure fields with an $SU(1,1)$ integer spin and a
half-integer spin, the discussion is not exactly the same as that
in this paper that contains three structure fields with
$SU(1,1)$ integer and half-integer spins.
The mixing structures of higgses and the others are derived by two
structure fields with an $SU(1,1)$ integer spin.
For higgses, the doublet-triplet mass splitting can be realized
without fine-tuning, which has been discussed
in Refs.~\cite{Inoue:2000ia,Yamatsu:2007} as mentioned above.
The Yukawa coupling structures in ``MSSM'' have already been discussed in
Ref.~\cite{Yamatsu:2007}. When the mixing structures of down-type quarks
and charged 
leptons include $SU(5)$ breaking effects, we will see that the GUT
relation for the Yukawa coupling structures of down-type quarks and 
charged lepton is avoided.
We will discuss the $\mu$-term, although the generation of the
$\mu$-term has been discussed in Ref.~\cite{Yamatsu:2008}, where the
matter content of singlets and the scalar potential is different. 
We will discuss that special weight assignments of
$SU(1,1)$ allow $R$-parity to remain even after the $SU(1,1)$ breaking,
where it was first pointed out that $\hat{L}\hat{L}\hat{E}^c$, 
$\hat{Q}\hat{L}\hat{D}^c$, $\hat{D}^c\hat{D}^c\hat{U}^c$ are absent in
Refs.~\cite{Inoue:1994qz,Yamatsu:2007}, and  $\hat{H}_u\hat{H}_d$ is also 
absent in Ref.~\cite{Yamatsu:2008} because all fields have the same sign
of weight. 
An article \cite{Inoue:2003qi} suggested that a $G_{\rm SM}\times SU(1,1)$
model with particular matter content allows only Type-II seesaw mechanism
\cite{Magg:1980ut,Lazarides:1980nt,Mohapatra:1980yp} to generate
neutrino masses. 
In general, not only Type-II seesaw mechanism but also Type-I and
Type-III seesaw mechanisms are allowed, where the $SU(1,1)$ weight
assignments are severely constrained. 
We will see that the proton decay via colored higgses is naturally
suppressed since the colored higgses have Dirac mass terms.
Note that this idea has already been discussed at least in the context of 
an orbifold GUT model based on extra dimension $S_1/Z_2\times Z_2$ in
Ref.~\cite{Hall:2001pg}, where any colored higgs has a Dirac mass term
by using a non-trivial boundary condition.

This paper is organized as follows. 
In Sec.~\ref{Sec:Setup}, we first set up our model.
In Sec.~\ref{Sec:SGG}, we discuss spontaneous generation of generations
to produce three chiral generations of quarks and leptons and one
generation  of higgses by using one structure field with a half-integer
spin of $SU(1,1)$ and two structure fields with integer spins as
proposed in Ref.~\cite{Yamatsu:2012}.
In particular, we find that the colored higgses can disappear without
fine-tuning.
In Sec.~\ref{Sec:Yukawa}, we see the structure of the Yukawa couplings,
especially how to realize the difference of the Yukawa coupling
matrices between the down-type quarks and charged leptons.
In Sec.~\ref{Sec:mu-term}, we discuss how to generate the effective
$\mu$-term of higgses.
In Sec.~\ref{Sec:BLBreaking}, we discuss the baryon and/or lepton number
violation including $R$-parity, neutrino masses, and proton decay.
We see that the proton decay derived from the colored higgses is
highly suppressed. Section~\ref{Sec:Summary} is devoted to a summary and
discussion.

\section{Setup of an $SU(5)\times SU(1,1)$ model}
\label{Sec:Setup}

We construct an $SU(5)$ SUSY GUT model with horizontal symmetry
$SU(1,1)$ that contains vectorlike matter content. We introduce the
matter fields 
\begin{align}
&\hat{F}_{{\bf 10}},\
\hat{F}_{{\bf 10}}^{\prime},\
\hat{G}_{{\bf 5}^*},\
\hat{G}_{{\bf 5}^*}^{\prime},\
\hat{H}_{u{\bf 5}},\
\hat{H}_{d{\bf 5}^*},\
\hat{S}_{\bf 1},\
\hat{R}_{\bf 1},\
\{\hat{N}_{\bf 1},\
\hat{T}_{\bf 15},\
\hat{A}_{\bf 24}\},\\
&\hat{F}_{{\bf 10}^*}^c,\
\hat{F}_{{\bf 10}^*}^{\prime c},\
\hat{G}_{{\bf 5}}^c,\
\hat{G}_{{\bf 5}}^{\prime c},\
\hat{H}_{u{\bf 5}^*}^c,\
\hat{H}_{d{\bf 5}}^c,\
\hat{S}_{\bf 1}^c,\
\hat{R}_{\bf 1}^c,\
\{\hat{N}_{\bf 1}^c,\
\hat{T}_{{\bf 15}^*}^c,\
\hat{A}_{\bf 24}^c\},
\end{align}
where the bold subscripts stand for the representations in $SU(5)$.
Since a pair of the fields in the curly brackets
$\{\hat{N}_{\bf 1}, \hat{T}_{\bf 15}, \hat{A}_{\bf 24}\}$ and
$\{\hat{N}_{\bf 1}^c, \hat{T}_{{\bf 15}^*}^c, \hat{A}_{\bf 24}^c\}$ 
are necessary to generate nonzero neutrino masses, we introduce 
one pair of them and in Sec.~\ref{Sec:BLBreaking} we will see 
which fields are compatible with the $SU(1,1)$ weight assignment
constrained by other requirements, such as to generate three chiral
generations of quarks and leptons and one chiral generations of higgses,
to allow Yukawa couplings between quarks and leptons and higgses. 
The quantum numbers of $SU(5)\times SU(1,1)$ and $R$-parity are
summarized in Table~\ref{tab:matter}.  
We define the values $q_\alpha$ and $q_\beta$ as 
\begin{align}
q_\alpha:=\alpha'-\alpha,\ \ \
q_\beta:=\beta'-\beta,
\end{align}
where $\alpha$, $\beta$, etc. are $SU(1,1)$ weights.
We choose the values $q_\alpha$ and $q_\beta$ to be positive
half-integers. See Ref.~\cite{Yamatsu:2012} in detail for the notation
and convention.

\begin{table}[t]
\begin{center}
\begin{tabular}{|c|cccc|cc|cc|ccc|}\hline
Field    &$\hat{F}_{\bf 10}$&$\hat{F}_{\bf 10}'$&
$\hat{G}_{{\bf 5}^*}$&$\hat{G}_{{\bf 5}^*}'$&
$\hat{H}_{u{\bf 5}}$&$\hat{H}_{d{\bf 5}^*}$&
$\hat{S}_{\bf 1}$&$\hat{R}_{\bf 1}$
&$\hat{N}_{\bf 1}$&$\hat{T}_{\bf 15}$&$\hat{A}_{\bf 24}$
\\\hline\hline
$SU(5)$  &${\bf 10}$&${\bf 10}$&${\bf 5}^*$&${\bf 5}^*$&
${\bf 5}$&${\bf 5}^*$&${\bf 1}$&${\bf 1}$
&${\bf 1}$&${\bf 15}$&${\bf 24}$\\
$SU(1,1)$&$+\alpha$&$+\alpha'$&$+\beta$&$+\beta'$&
$-\gamma$&$-\delta$&$+\eta$&$+\lambda$&$+\xi$&$-\tau$&$+\zeta$\\
($R$-parity)&$-$&$-$&$-$&$-$&
$+$&$+$&$+$&$+$&$-$&$+$&$-$\\
\hline
\end{tabular}
\end{center}
\caption{The quantum numbers of matter fields in the 
$SU(5)\times SU(1,1)$ model are given in the table, and the model has
also their conjugate fields. 
The Greek letters of the $SU(1,1)$ row represent the highest or
lowest eigenvalues of $SU(1,1)$ weights. The negative value is
the highest weight and the positive value is  the lowest weight of
$SU(1,1)$.
\label{tab:matter}}
\end{table}

We also introduce the structure fields
\begin{align}
\hat\Phi_{{\bf 1}},\ \hat\Phi_{{\bf 24}}',\  \hat\Psi_{{\bf 1}/{\bf 24}},
\end{align}
where the $SU(1,1)$ spins of 
$\hat\Phi_{{\bf 1}}$, $\hat\Phi_{{\bf 24}}'$, and 
$\hat\Psi_{{\bf 1}/{\bf 24}}$ are $S$, $S'$, and $S''$, respectively.
This is summarized in Table~\ref{tab:structure}.
The subscript of $\hat\Psi_{{\bf 1}/{\bf 24}}$ represents two options for
the $SU(5)$ representations.

We assume that the gauge group $SU(5)\times SU(1,1)$ is spontaneously
broken to $G_{\rm SM}$ via the following nonvanishing VEVs of the
structure fields
\begin{align}
\langle\hat{\Phi}_{\bf 1}\rangle=\langle\phi_{0}\rangle,
\hspace{1em}
\langle\hat{\Phi}_{\bf 24}'\rangle=\langle \phi_{+1}'\rangle,
\hspace{1em}
\langle\hat{\Psi}_{{\bf 1}/{\bf 24}}\rangle=\langle \psi_{-3/2}\rangle,
\label{Structure-VEVs}
\end{align}
where the subscripts of $\langle\phi_{0}\rangle$,
$\langle \phi_{+1}'\rangle$ and $\langle \psi_{-3/2}\rangle$
stand for the eigenvalues of the third component generator of $SU(1,1)$.
In the next section, we will find that the $SU(1,1)$ spins must satisfy
$S=S'< S''$, and to realize three generations of quarks and leptons and
one generations of higgses, the minimal choice is $S=S'=1$ and $S''=3/2$.
We will also find that the VEV of $\hat{\Phi}_{\bf 24}'$ plays essential
roles for decomposing the doublet and triplet higgses and making
difference between the Yukawa coupling constants of the down-type quarks
and charged leptons. The Clebsch-Gordan coefficients (CGCs) of $SU(5)$ are
shown in Ref.~\cite{Slansky:1981yr,Koh:1982zz,DelOlmo:1987bc}. 
The CGCs of $SU(1,1)$ are found in Ref.~\cite{Yamatsu:2012}. 

\begin{table}[t]
\begin{center}
\begin{tabular}{|c|ccc|}\hline
Field    &$\hat{\Phi}_{\bf 1}$&$\hat{\Phi}_{\bf 24}'$&
$\hat{\Psi}_{{\bf 1}/{\bf 24}}$\\\hline\hline
$SU(5)$  &${\bf 1}$&${\bf 24}$&${\bf 1}$ or ${\bf 24}$\\
$SU(1,1)$&$S$&$S'$&$S''$\\
($R$-parity)&$+$&$+$&$+$\\
\hline
\end{tabular}
\end{center}
\caption{The quantum numbers of structure fields in the 
$SU(5)\times SU(1,1)$ model are given in the table.
\label{tab:structure}}
\end{table}

We describe other assumptions as follows. 
The gauge kinetic function of the $SU(1,1)$ vector superfield and the
K\"ahler potential of the structure fields have positive definite
metrics at a vacuum. 
(Note that to realize this situation, at least one nonrenormalizable
term must have larger effects for metrics of the SU(1,1) gauge and the
structure fields than their renormalizable terms in this model.)
The Lagrangian in the matter field sector
including the coupling terms between matter fields and structure fields
contains only renormalizable terms, and 
non-renormalizable terms in superpotential are induced by the
process of decoupling the heavy fields.
The correction for the K\"ahler potential of matter fields and the
gauge kinetic function of the $SU(5)$ gauge fields is negligible.
After the chiral fields are generated via the spontaneous generation of
generations 
\cite{Inoue:1994qz,Inoue:2000ia,Yamatsu:2007,Yamatsu:2012}, 
the effect from the $SU(1,1)$ gauge bosons and the structure fields
is negligible for the chiral matter fields at low energy.
Only the structure fields have large VEVs and
the matter fields have smaller VEVs compared to those of
the structure fields because of maintaining the structures of the
horizontal symmetry; e.g., the VEVs of the structure fields are
GUT-scale mass  $M_{\rm GUT}\simeq O(10^{16})$ GeV and the 
VEVs of the matter fields are $m_{\rm SUSY}\simeq O(10^{3})$ GeV.
Some $SU(1,1)$ singlet superfields break SUSY in a hidden
sector, SUSY breaking does not affect $SU(1,1)$ symmetry, and 
soft SUSY breaking terms for matter fields are generated at GUT scale
$M_{\rm GUT}\sim O(10^{16})$ GeV in a visible sector, where the soft
SUSY breaking masses are $O(m_{\rm SUSY})\sim O(10^3)$ GeV. 
To discuss the D-flatness condition of the $SU(1,1)$ group, we would
have to consider the full potential of the model, including all
structure fields because the D-flatness condition depends
on the K\"ahler potential of the structure field. 
We therefore neglect this effect in this paper. 

The number of soft SUSY breaking terms are determined by the number of
the superpotential terms. It is impossible to give explicit forms of the
soft SUSY braking terms before we discuss the superpotential. Here we
mention the pattern of the soft SUSY breaking terms. 
Under the above assumption, $SU(1,1)$ symmetry restricts the structures
of the soft SUSY breaking terms up to renormalizable terms;
each trilinear scalar term, so-called $A$-term, is  exactly
proportional to the corresponding Yukawa coupling term in the
superpotential; soft scalar masses are generation-independent.
Note that the pattern of the soft SUSY terms can change when we take
into account of the higher order terms derived from the
non-renormalizable terms of the K\"ahler potential and the
superpotential.

\section{Realization of chiral generations}
\label{Sec:SGG}

In this section, we consider how to provide three chiral generations of
quarks and leptons, one chiral generation of up- and down-type doublet
higgses, and no chiral generations of the others shown in 
Table~\ref{tab:matter} by using three structure fields shown in 
Table~\ref{tab:structure}.
We use the methods developed in Ref.~\cite{Yamatsu:2012}.

\subsection{Three chiral generations of quarks and leptons}

We discuss how to produce three chiral generations of quarks and
leptons. The superpotential of the quark and lepton superfields coupling
to the structure fields is given by 
\begin{align}
W=&M_f\hat{F}_{\bf 10}\hat{F}_{{\bf 10}^*}^c
+M_f'\hat{F}_{\bf 10}^{\prime}\hat{F}_{{\bf 10}^*}^{\prime c}
+x_f\hat{F}_{\bf 10}\hat{F}_{{\bf 10}^*}^c\hat\Phi_{{\bf 1}}
+x_f'\hat{F}_{\bf 10}^{\prime}\hat{F}_{{\bf 10}^*}^{\prime c}
\hat\Phi_{{\bf 1}}\nonumber\\
&+z_f\hat{F}_{\bf 10}\hat{F}_{{\bf 10}^*}^c
\hat\Phi_{{\bf 24}}'
+z_f'\hat{F}_{{\bf 10}}^{\prime}
\hat{F}_{{\bf 10}^*}^{\prime c}
\hat\Phi_{{\bf 24}}'
+w_f\hat{F}_{{\bf 10}}^{\prime}
\hat{F}_{{\bf 10}^*}^c
\hat\Psi_{{\bf 1}/{\bf 24}}
+w_f'\hat{F}_{\bf 10}
\hat{F}_{{\bf 10}^*}^{\prime c}
\hat\Psi_{{\bf 1}/{\bf 24}}\nonumber\\
&
+M_g\hat{G}_{{\bf 5}^*}\hat{G}_{{\bf 5}}^c
+M_g'\hat{G}_{{\bf 5}^*}^{\prime}\hat{G}_{{\bf 5}}^{\prime c}
+x_g\hat{G}_{{\bf 5}^*}\hat{G}_{{\bf 5}}^c\hat\Phi_{{\bf 1}}
+x_g'\hat{G}_{{\bf 5}^*}^{\prime}\hat{G}_{{\bf 5}}^{\prime c}
\hat\Phi_{{\bf 1}}\nonumber\\
&+z_g\hat{G}_{{\bf 5}^*}\hat{G}_{{\bf 5}}^c
\hat\Phi_{{\bf 24}}'
+z_g'\hat{G}_{{\bf 5}^*}^{\prime}
\hat{G}_{{\bf 5}}^{\prime c}\hat\Phi_{{\bf 24}}'
+w_g\hat{G}_{{\bf 5}^*}^{\prime}\hat{G}_{{\bf 5}}^c
\hat\Psi_{{\bf 1}/{\bf 24}}
+w_g'\hat{G}_{{\bf 5}^*}\hat{G}_{{\bf 5}}^{\prime c}
\hat\Psi_{{\bf 1}/{\bf 24}},
\label{W:quarks-leptons}
\end{align}
where $M$s are mass parameters, $x$s, $z$s, and $w$s are dimensionless
coupling constants.
We assume that the massless chiral fields are realized as linear
combinations of the components of $\hat{F}_{\bf 10}$, 
$\hat{F}_{\bf 10}'$, $\hat{G}_{{\bf 5}^*}$, 
and $\hat{G}_{{\bf 5}^*}'$ in the manner
\begin{align}
&\hat{f}_{\alpha+i}=\sum_{n=0}^{2}\hat{f}_nU_{n,i}^{f}
+[\mbox{massive modes}],\ \ \ 
\hat{f}_{\alpha'+i}^{\prime}=\sum_{n=0}^{2}\hat{f}_nU_{n,i}^{f'}
+[\mbox{massive modes}],
\end{align}
where $\hat{F}_{\bf 10}$ and $\hat{F}_{\bf 10}'$ contain
$\hat{Q}$, $\hat{U}^c$, and $\hat{E}^c$,
and $\hat{G}_{{\bf 5}^*}$ and $\hat{G}_{{\bf 5}^*}'$ contain
$\hat{D}^c$ and $\hat{L}$. $f$ stands for $q$, $u^c$, $e^c$, $d^c$, and
$\ell$. For $f=d^c,$ $\ell$, $\alpha$ should be replaced by $\beta$.

We solve the massless condition by using the mass term of the
superpotential in Eq.~(\ref{W:quarks-leptons}) for the matter field
$\hat{F}_{\bf 10}$.  For this calculation, there is no difference between
the matter fields $\hat{F}_{\bf 10}$ and $\hat{G}_{{\bf 5}^*}$ 
except the coupling constants and some CGCs. 
By substituting the nonvanishing VEVs of the structure fields in
Eq.~(\ref{Structure-VEVs}) into the superpotential term in
Eq.~(\ref{W:quarks-leptons}), we have the mass term 
\begin{align}
\left.W\right|_{\Phi=\langle\Phi\rangle}
&=M_f\hat{F}\hat{F}^c
+M_f'\hat{F}^{\prime}\hat{F}^{\prime c}
+x_f\hat{F}\hat{F}^c\langle\hat\Phi_{{\bf 1}}\rangle
+x_f'\hat{F}^{\prime}\hat{F}^{\prime c}
\langle\hat\Phi_{{\bf 1}}\rangle\nonumber\\
&+Y_fz_f\hat{F}\hat{F}^c\langle\hat\Phi_{{\bf 24}}'\rangle
+Y_fz_f'\hat{F}^{\prime}\hat{F}^{\prime c}
\langle\hat\Phi_{{\bf 24}}'\rangle
+\tilde{Y}_fw_f\hat{F}^{\prime}\hat{F}^c
\langle\hat\Psi_{{\bf 1}/{\bf 24}}\rangle
+\tilde{Y}_fw_f'\hat{F}\hat{F}^{\prime c}
\langle\hat\Psi_{{\bf 1}/{\bf 24}}\rangle\nonumber\\
&\hspace{-3em}
=\sum_{i=0}^\infty
\bigg[
\left(M_f(-1)^i+x_f\langle\phi_{0}\rangle 
D_{i,i}^{\alpha,\alpha,S}\right)
\hat{f}_{\alpha+i}\hat{f}_{-\alpha-i}^{c}
+\left(M_f'(-1)^i
+x'\langle\phi_{0}\rangle D_{i,i}^{\alpha',\alpha',S}\right)
\hat{f}_{\alpha'+i}^{\prime}\hat{f}_{-\alpha'-i}^{\prime c}\nonumber\\
&\hspace{-2em} 
+Y_fz_f\langle\phi_{+1}'\rangle D_{i+1,i}^{\alpha,\alpha,S'}
\hat{f}_{\alpha+i}\hat{f}_{-\alpha-i-1}^{c}
+Y_fz_f'\langle\phi_{+1}'\rangle D_{i+1,i}^{\alpha',\alpha',S'}
\hat{f}_{\alpha'+i}^{\prime}\hat{f}_{-\alpha'-i-1}^{\prime c}\nonumber\\
&\hspace{-2em} 
+\tilde{Y}_fw_f\langle\psi_{-3/2}\rangle
D_{i,i+3/2-q_\alpha}^{\alpha,\alpha',S''}
\hat{f}_{\alpha'+i+3/2-q_\alpha}^{\prime}\hat{f}_{-\alpha-i}^{c}
+\tilde{Y}_fw_f'\langle\psi_{-3/2}\rangle
D_{i,i+3/2+q_\alpha}^{\alpha',\alpha,S''}
\hat{f}_{\alpha+i+3/2+q_\alpha}\hat{f}_{-\alpha'-i}^{\prime c}\bigg]
\nonumber\\
&\hspace{-3em}
=\sum_{n=0}\sum_{i=0}^\infty
\bigg[\hat{f}_n\bigg\{
\left(M_f(-1)^i+x_f\langle\phi_{0}\rangle 
D_{i,i}^{\alpha,\alpha,S}\right)U_{n,i}^f
+Y_fz_f\langle\phi_{+1}'\rangle D_{i,i-1}^{\alpha,\alpha,S'}
U_{n,i-1}^f\nonumber\\
&
+\tilde{Y}_fw_f\langle\psi_{-3/2}\rangle
D_{i,i+3/2-q_\alpha}^{\alpha,\alpha',S''}
U_{n,i+3/2-q_\alpha}^{f'}
\bigg\}\hat{f}_{-\alpha-i}^{c}\nonumber\\
&+\hat{f}_n\bigg\{
\left(M_f'(-1)^i
+x_f'\langle\phi_{0}\rangle D_{i,i}^{\alpha',\alpha',S}\right)
U_{n,i}^{f'}
+Y_fz_f'\langle\phi_{+1}'\rangle D_{i+1,i}^{\alpha',\alpha',S'}
U_{n,i-1}^{f'}\nonumber\\
&+\tilde{Y}_fw_f'\langle\psi_{-3/2}\rangle
D_{i,i+3/2+q_\alpha}^{\alpha',\alpha,S''}
U_{n,i+3/2+q_\alpha}^f
\bigg\}\hat{f}_{-\alpha'-i}^{\prime c}\bigg]+[\mbox{massive modes}],
\label{W:FF-mass}
\end{align}
where $Y_f$ is a $U(1)_Y$ charge shown in Table~\ref{tab:SM-reps}, and
$\tilde{Y}_f$ is equal to one for $\hat{\Psi}_{\bf 1}$ and is equal to
$Y_f$ for $\hat{\Psi}_{\bf 24}$. 
$D_{j,i}^{\beta,\alpha,S}$ ($i,j=0,1,2,\cdots$) is a CGC of $SU(1,1)$
given in Ref.~\cite{Yamatsu:2012}; for $S\geq |-i+j-\alpha+\beta|$, the
CGC is nonzero; otherwise, the CGC is zero.
The emergence of the massless modes $\hat{f}_n$ requires that the
coupling between the massless modes $\hat{f}_n$ and the massive modes
$\hat{f}_{-\alpha-i}^c$ and $\hat{f}_{-\alpha-i}^{\prime c}$ for any $i$
must vanish simultaneously:
\begin{align}
&X_iU_{n,i}^f+Y_iU_{n,i-1}^f+Z_iU_{n,i+3/2-q_\alpha}^{f'}=0,\\
&X_i'U_{n,i}^{f'}+Y_i'U_{n,i-1}^{f'}+Z_i'U_{n,i+3/2+q_\alpha}^f=0,
\end{align}
where we defined 
\begin{align}
&X_i:=M_f(-1)^i+x_f\langle\phi_{0}\rangle
D_{i,i}^{\alpha,\alpha,S},\ 
Y_i:=Y_fz_f\langle\phi_{+1}'\rangle D_{i,i-1}^{\alpha,\alpha,S'},\ 
Z_i:=\tilde{Y}_fw_f\langle\psi_{-3/2}\rangle
D_{i,i+3/2-q_\alpha}^{\alpha,\alpha',S''},\nonumber\\
&X_i':=M_f'(-1)^i
+x_f'\langle\phi_{0}\rangle D_{i,i}^{\alpha',\alpha',S},\ 
Y_i':=Y_fz_f'\langle\phi_{+1}'\rangle D_{i+1,i}^{\alpha',\alpha',S'},\ 
Z_i':=\tilde{Y}_fw_f'\langle\psi_{-3/2}\rangle
D_{i,i+3/2+q_\alpha}^{\alpha',\alpha,S''}.
\end{align}
These lead to the relation among the mixing coefficients $U_{n,i}^f$
and $U_{n,i}^{f'}$, respectively:
\begin{align}
U_{n,i+3}^f=&
\frac{X_{i+3/2-q_\alpha}'}{Z_{i+3/2-q_\alpha}'}\frac{X_i}{Z_i}
U_{n,i}^f
+\left(\frac{X_{i+3/2-q_\alpha}'}{Z_{i+3/2-q_\alpha}'}\frac{Y_i}{Z_i}
+\frac{Y_{i+3/2-q_\alpha}'}{Z_{i+3/2-q_\alpha}'}\frac{X_{i-1}}{Z_{i-1}}
\right)U_{n,i-1}^f
+\frac{Y_{i+3/2-q_\alpha}'}{Z_{i+3/2-q_\alpha}'}\frac{Y_{i-1}}{Z_{i-1}}
U_{n,i-2}^f,
\label{MC:quarks-leptons-1}
\\
U_{n,i+3}^{f'}=&
\frac{X_{i+3/2+q_\alpha}}{Z_{i+3/2+q_\alpha}}\frac{X_i'}{Z_i'}
U_{n,i}^{f'}
+\left(\frac{X_{i+3/2+q_\alpha}}{Z_{i+3/2+q_\alpha}}\frac{Y_i'}{Z_i'}
+\frac{Y_{i+3/2+q_\alpha}}{Z_{i+3/2+q_\alpha}}\frac{X_{i-1}'}{Z_{i-1}'}
\right)U_{n,i-1}^{f'}
+\frac{Y_{i+3/2+q_\alpha}}{Z_{i+3/2+q_\alpha}}\frac{Y_{i-1}'}{Z_{i-1}'}
U_{n,i-2}^{f'},
\label{MC:quarks-leptons-2}
\end{align}
where for $i<0$, $U_{n,i}^f=U_{n,i}^{f'}=0$. 

\begin{table}[t]
\begin{center}
\begin{tabular}{|c|ccccc|cccc|}\hline
Field    &$\hat{Q}^{(\prime)}$&$\hat{U}^{c(\prime)}$&
$\hat{D}^{c(\prime)}$&$\hat{L}^{(\prime)}$&$\hat{E}^{c(\prime)}$
&$\hat{H}_u$&$\hat{H}_d$&$\hat{T}_u$&$\hat{T}_d$\\\hline\hline
$SU(3)_C$&${\bf 3}$&${\bf 3}^*$&${\bf 3}^*$&${\bf 1}$&${\bf 1}$
&${\bf 1}$&${\bf 1}$&${\bf 3}$&${\bf 3}^*$\\
$SU(2)_L$&${\bf 2}$&${\bf 1}$&${\bf 1}$&${\bf 2}$&${\bf 1}$
&${\bf 2}$&${\bf 2}$&${\bf 1}$&${\bf 1}$\\
$U(1)_Y$&$+1/6$&$-2/3$&$+1/3$&$-1/2$&$+1$
&$+1/2$&$-1/2$&$-1/3$&$+1/3$\\\hline
\end{tabular}
\end{center}
\caption{The quantum numbers of 
$G_{SM}=SU(3)_C\times SU(2)_L\times U(1)_Y$ 
for matter fields in the $SU(5)\times SU(1,1)$ model are given in the
table, and their conjugate fields have conjugate representations.
\label{tab:SM-reps}}
\end{table}

\begin{figure}
\centering
\includegraphics[totalheight=6cm]{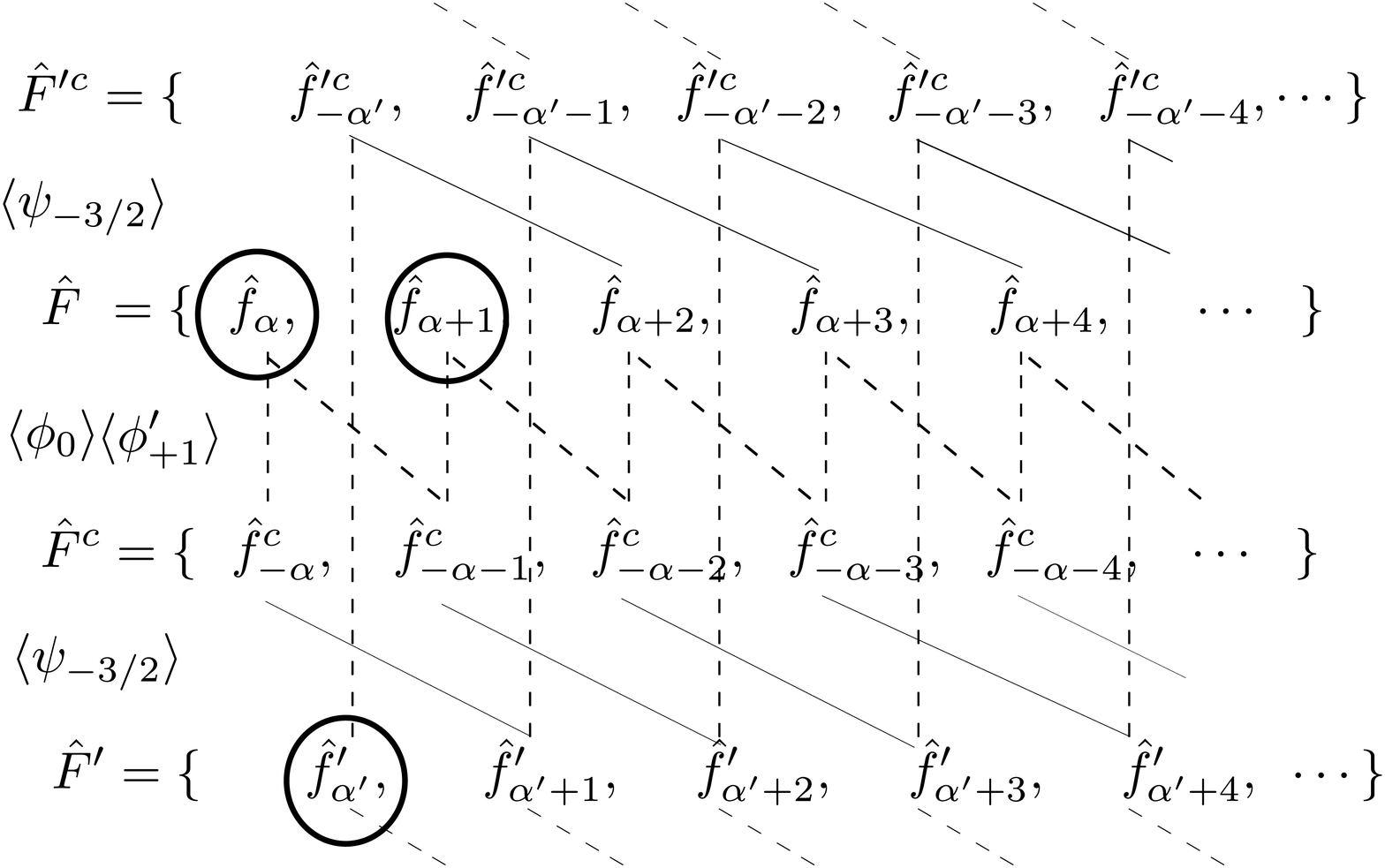}
\caption{Type-I $q_\alpha=1/2$ for the three massless
generation of quarks and leptons in Eq.~(\ref{W:FF-mass}):
a massless mode $\hat{f}_0$ is realized as certain linear combinations of
the components $\hat{f}_{\alpha+k}$ and $\hat{f}'_{\alpha'+1+k}$
$(k=0,1,2,\cdots)$, 
and the constructional element of the massless mode is determined  by its
mixing coefficients $U_{0,k}^f$ and $U_{0,k+1}^{f'}$
given in Eqs.~(\ref{MC:quarks-leptons-1}) and
(\ref{MC:quarks-leptons-2}) and Type-I initial condition;
a massless mode $\hat{f}_1$ is realized as certain linear combinations of
the components $\hat{f}_{\alpha+k+1}$ and $\hat{f}'_{\alpha'+2+k}$
$(k=0,1,2,\cdots)$, 
and the constructional element of the massless mode is determined  by its
mixing coefficients $U_{1,k+1}^f$ and $U_{1,k+2}^{f'}$
given in Eqs.~(\ref{MC:quarks-leptons-1}) and
(\ref{MC:quarks-leptons-2}) and Type-I initial condition;
a massless mode $\hat{f}_2$ is realized as certain linear combinations of
the components $\hat{f}_{\alpha+k+2}$ and $\hat{f}'_{\alpha'+k}$
$(k=0,1,2,\cdots)$, 
and the constructional element of the massless mode is determined  by its
mixing coefficients $U_{2,k+2}^f$ and $U_{2,k}^{f'}$
given in Eqs.~(\ref{MC:quarks-leptons-1}) and
(\ref{MC:quarks-leptons-2}) and Type-I initial condition.
The components of the matter fields $\hat{F}$, $\hat{F}^c$,
$\hat{F}'$ and $\hat{F}^{\prime c}$
connected by solid and dashed lines have a mass term that 
comes from the VEVs $\langle\psi_{-3/2}\rangle$ and,
$\langle\phi_{0}\rangle$ and $\langle\phi_{+1}'\rangle$, respectively.
The mass term of the solid line that comes from the VEV
$\langle\psi_{-3/2}\rangle$ dominantly contributes to whether 
massless modes appear or not.
(It does not always dominantly contribute to small components
of the mixing coefficients.)
The figure wraps from bottom to top.
A component surrounded by a circle is a main element of each massless
chiral mode when the mass terms of the solid line dominantly
contribute to small components of the mixing coefficients.
}
\label{fig:SGG-new-p=3/2-T1}
\end{figure}

\begin{figure}
\centering
\includegraphics[totalheight=6cm]{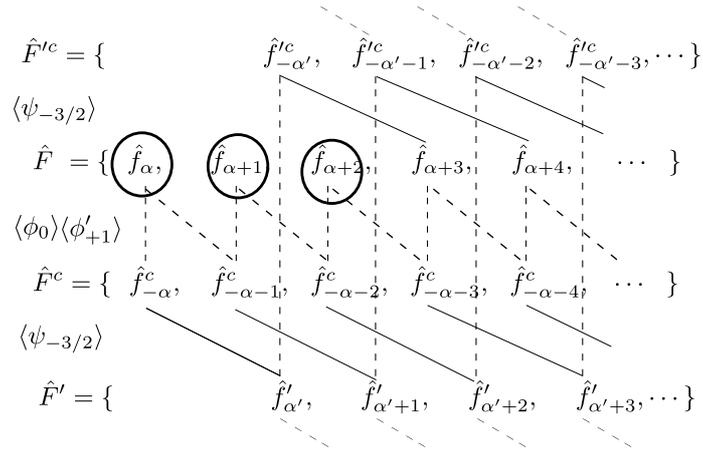}
\caption{
Type-II $q_\alpha=3/2$ for the three massless
generation of quarks and leptons in Eq.~(\ref{W:FF-mass}):
each massless mode $\hat{f}_n$ is realized as a certain linear combination
of the components $\hat{f}_{\alpha+k+n}$ and $\hat{f}'_{\alpha'+k+n}$
$(k=0,1,2,\cdots)$, 
and the constructional element of its massless mode is determined by its
mixing coefficients $U_{n,k+n}^f$ and $U_{n,k+n}^{f'}$
given in Eqs.~(\ref{MC:quarks-leptons-1}) and
(\ref{MC:quarks-leptons-2}) and Type-II initial condition.
The explanation of the circle and lines is given 
in Fig.~\ref{fig:SGG-new-p=3/2-T1}.}
\label{fig:SGG-new-p=3/2-T2}
\end{figure}

\begin{figure}
\centering
\includegraphics[totalheight=6cm]{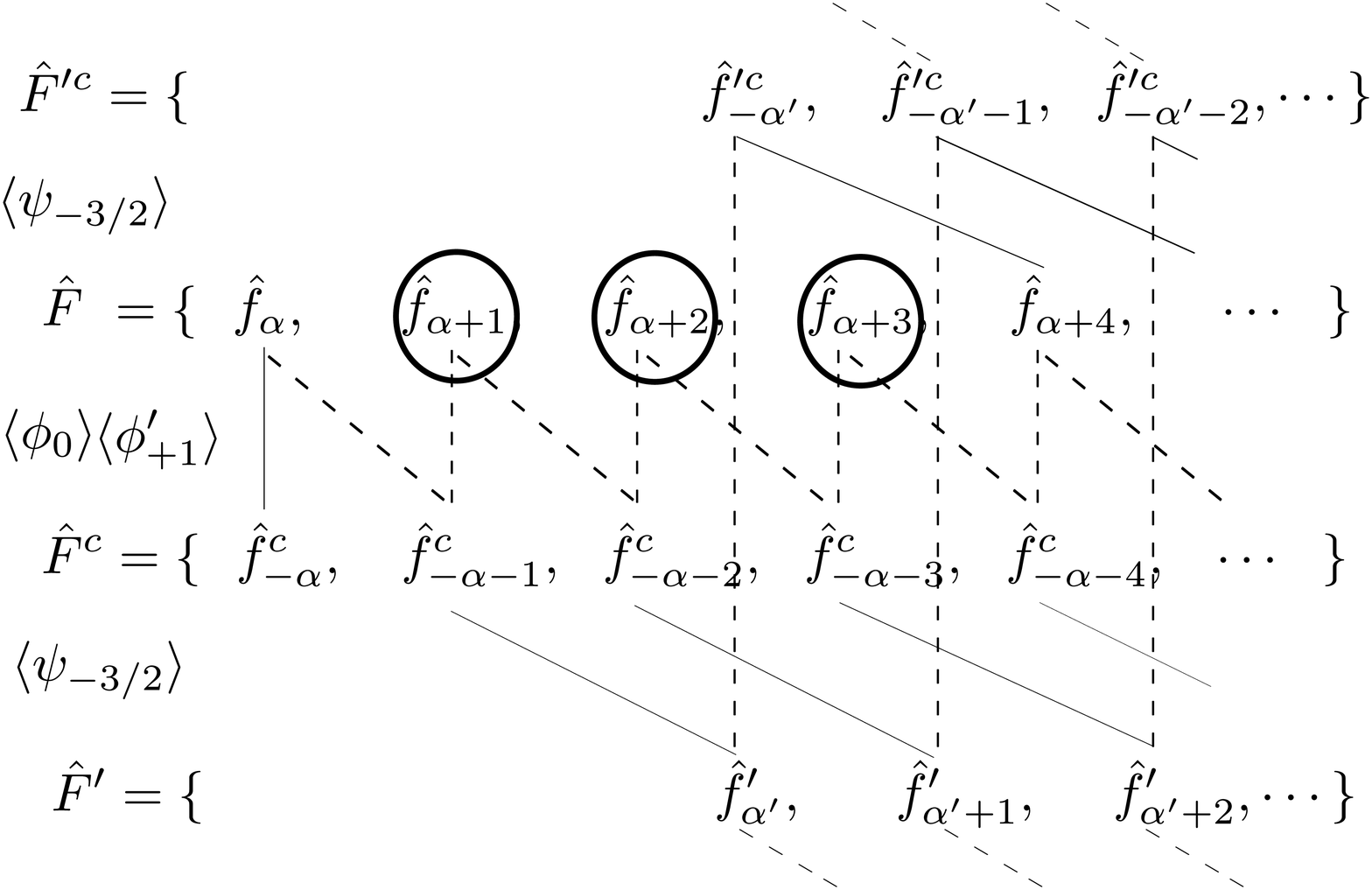}
\caption{An example of Type-III $q_\alpha=5/2$ for the three massless
generation of quarks and leptons in Eq.~(\ref{W:FF-mass}):
each massless mode $\hat{f}_n$ is realized as a certain linear combination
of the components $\hat{f}_{\alpha+k+n+1}$ and $\hat{f}'_{\alpha'+k+n}$
$(k=0,1,2,\cdots)$, 
and the constructional element of its massless mode is determined by its
mixing coefficients $U_{n,k+n+1}^f$ and $U_{n,k+n}^{f'}$
given in Eqs.~(\ref{MC:quarks-leptons-1}) and
(\ref{MC:quarks-leptons-2}) and Type-III initial condition.
The explanation of the circle and lines is given 
in Fig.~\ref{fig:SGG-new-p=3/2-T1}.}
\label{fig:SGG-new-p=3/2-T3}
\end{figure}

The recursion equations determine the mixing coefficients $U_{n,i}^f$
and $U_{n,i}^{f'}$ for any $i$.
The two initial condition sets of the mixing coefficients $U_{n,i}^f$
and $U_{n,i}^{f'}$ are also dependent upon each other.
As in Ref.~\cite{Yamatsu:2012}, Sec.~3.2, the relation between two
initial condition sets can be classified into three conditions: Type-I,
$q_\alpha<3/2$ shown in Fig.~\ref{fig:SGG-new-p=3/2-T1}; 
Type-II, $q_\alpha=3/2$ shown in Fig.~\ref{fig:SGG-new-p=3/2-T2};
and Type-III, $q_\alpha>3/2$ shown in Fig.~\ref{fig:SGG-new-p=3/2-T3}.
Each condition leads to different mixing coefficients. When we
calculate the Yukawa coupling constants, we need to have their detailed
information. In this paper, we will not analyze the Yukawa couplings in
detail. We will just show the difference
between the Yukawa couplings of down-type quarks and charged leptons in
Sec.~\ref{Sec:Yukawa}.

We need to consider the normalizable condition of the mixing
coefficients $U_{n,i}^{f}$ and $U_{n,j}^{f'}$.
As in Ref.~\cite{Yamatsu:2012}, Sec.~3.2, when the $SU(1,1)$ spins
satisfy the condition $S''>S,S'$, 
their normalizable conditions are always satisfied regardless of the
coupling constants and the value of the VEVs. Thus, three massless modes
$\hat{f}_n$ $(n=0,1,2)$ appear at low energy. 
Note that when we make Figs.~\ref{fig:SGG-new-p=3/2-T1},
\ref{fig:SGG-new-p=3/2-T2} and \ref{fig:SGG-new-p=3/2-T3},
we have already assumed $S''>S,S'$.

One may notice that if the quark and lepton superfield 
$\hat{G}_{{\bf 5}^*}^{(\prime)}$ and the conjugate superfield of up-type
higgs $\hat{H}_{u{\bf 5}^*}^c$  mix with each other, then this
discussion is ruined.  
Thus, if this model does not include explicitly e.g. the $R$-parity
shown in Table~\ref{tab:matter} and \ref{tab:structure}, the $SU(1,1)$
weight 
content must satisfy $\gamma\not=\beta+[\mbox{integer or half-integer}]$.
In the model with $R$-parity shown in Table~\ref{tab:matter} and
\ref{tab:structure}, the quantum number of the quark and lepton
superfield  $\hat{G}_{{\bf 5}^*}^{(\prime)}$ is different from the
conjugate field of up-type higgs $\hat{H}_{u{\bf 5}^*}^c$, so there is
no such restriction.

\subsection{One chiral generation of higgses}
\label{Sec:SGG:Higgs}

We will see that the model can allow only one generation of up- and
down-type $SU(2)_L$ doublet higgses and prohibit any generation of 
the up- and down-type $SU(3)_C$ triplet higgses, so-called colored higgs,
at low energy without fine-tuning and unnatural parameter choices in the
sense of 't Hooft naturalness \cite{'tHooft:1980xb}. This is pointed out
in Refs.~\cite{Inoue:2000ia,Yamatsu:2007}.

We consider how to provide one chiral generations of higgses.
The superpotential for the matter and structure coupling  is
\begin{align}
W=&
M_{h_u}\hat{H}_{u{\bf 5}}\hat{H}_{u{\bf 5}^*}^c
+x_{h_u}\hat{H}_{u{\bf 5}}\hat{H}_{u{\bf 5}^*}^c\hat\Phi_{{\bf 1}}
+z_{h_u}\hat{H}_{u{\bf 5}}\hat{H}_{u{\bf 5}^*}^c\hat\Phi_{{\bf 24}}'
\nonumber\\
&
+M_{h_d}\hat{H}_{d{\bf 5}^*}\hat{H}_{d{\bf 5}}^c
+x_{h_d}\hat{H}_{d{\bf 5}^*}\hat{H}_{d{\bf 5}}^c\hat\Phi_{{\bf 1}}
+z_{h_d}\hat{H}_{d{\bf 5}^*}\hat{H}_{d{\bf 5}}^c\hat\Phi_{{\bf 24}}',
\label{W:higgses}
\end{align}
where $M$s are mass parameters, $x$s and $z$s are dimensionless coupling
constants.
We assume that one massless chiral generation of higgses is realized at
low energy as a linear combination of the components of 
$\hat{H}_{u{\bf 5}}$ and $\hat{H}_{d{\bf 5}^*}$ in the manner
\begin{align}
&\hat{h}_{-\gamma-i}=\hat{h}U_i^{h}
+[\mbox{massive modes}],
\end{align}
where $\hat{H}_{u{\bf 5}}$  contains $\hat{H}_u$ and
$\hat{T}_u$. $\hat{H}_{d{\bf 5}^*}$ contains $\hat{H}_d$ and
$\hat{T}_d$. $h$ represents $h_u$, $h_d$, $t_u$, and $t_d$.
For $h=h_d$ and $t_d$, $\gamma$ must be replaced by $\delta$.

We solve the massless condition by
using the mass term of the superpotential in Eq.~(\ref{W:higgses}).
The same as the quarks and lepton, we use generic notation $\hat{H}$ for
the higgs fields. By substituting the nonvanishing VEVs of the structure
fields in Eq.~(\ref{Structure-VEVs}) into the superpotential term in
Eq.~(\ref{W:higgses}), we have the mass term 
\begin{align}
\left.W\right|_{\Phi=\langle\Phi\rangle}
=&M_h\hat{H}\hat{H}^c
+x_h\hat{H}\hat{H}^c\langle\hat\Phi_{{\bf 1}}\rangle
+Y_hz_h\hat{H}\hat{H}^c\langle\hat\Phi_{{\bf 24}}'\rangle\nonumber\\
=&\sum_{i=0}^\infty
\bigg[
\left(M_h(-1)^i+x_h\langle\phi_{0}\rangle 
D_{i,i}^{\gamma,\gamma,S}\right)
\hat{h}_{-\gamma-i}\hat{h}_{\gamma+i}^{c}
+Y_hz_h\langle\phi_{+1}'\rangle D_{i+1,i}^{\gamma,\gamma,S'}
\hat{h}_{-\gamma-i-1}\hat{h}_{\gamma+i}^{c}
\nonumber\\
=&\sum_{i=0}^\infty
\hat{h}\bigg\{
\left(M_h(-1)^i+x_h\langle\phi_{0}\rangle 
D_{i,i}^{\gamma,\gamma,S}\right)U_{i}^h
+Y_hz_h\langle\phi_{+1}'\rangle D_{i+1,i}^{\gamma,\gamma,S'}
U_{i+1}^h
\bigg\}\hat{h}_{\gamma+i}^c\nonumber\\
&+[\mbox{massive modes}],
\label{W:higgses:vacuum}
\end{align}
where $Y_h$ is a $U(1)_Y$ charge shown in Table~\ref{tab:SM-reps}.

The massless mode $\hat{h}$ is extracted from the component
$\hat{h}_{-\gamma-i}$ of the matter field $\hat{H}$. The orthogonality
of the massless modes $\hat{h}$ to the massive modes
$\hat{h}_{\gamma+i}^c$ requires the coefficients $U_i^f$ to satisfy the
following recursion equation for any $i(\geq 0)$
\begin{align}
\left(M_h(-1)^i+x_h\langle\phi_{0}\rangle 
D_{i,i}^{\gamma,\gamma,S}\right)U_{i}^h
+Y_hz_h\langle\phi_{+1}'\rangle D_{i+1,i}^{\gamma,\gamma,S'}
U_{i+1}^h=0.
\end{align}
The relation of the mixing coefficients between $i$th and $i+1$th
components is
\begin{align}
U_{i+1}^h=
-\frac{M_h(-1)^i+x_h\langle\phi_{0}\rangle D_{i,i}^{\gamma,\gamma,S}}
{Y_hz_h\langle\phi_{+1}'\rangle D_{i+1,i}^{\gamma,\gamma,S'}}
U_i^h.
\label{Mixing-coefficients-h}
\end{align}

As is discussed in Ref.~\cite{Yamatsu:2012}, Sec.~3.1, we need to
consider a normalizable condition $\sum_{i=0}^\infty|U_{i}^h|<\infty$.
For $S=S'$, this leads to constraints for the values of the parameters and
the nonvanishing VEVs, where we will not consider the $SU(1,1)$ spins
satisfying $S<S'$ and $S>S'$ because the condition $S<S'$ provide one
chiral doublet and colored higgses and the condition $S>S'$ cannot
produce anything at low energy. By using the property of the CGC
$D_{i,j}^{\gamma,\gamma,S}$, for the large $i$ limit
Eq.~(\ref{Mixing-coefficients-h}) becomes
\begin{align}
\frac{U_{i+1}^h}{U_i^h}\sim
\frac{1}{Y_h}\frac{x_h\langle\phi_0\rangle}{z_h\langle\phi_{+1}'\rangle}
\sqrt{\frac{(S+1)!(S-1)!}{S!S!}},
\label{Mixing-coefficients-h-assy}
\end{align}
where we dropped the irrelevant term. To satisfy the normalizable
condition $\sum_{i=0}^\infty|U_{i}^h|<\infty$, the 
$|U_{i+1}^h/U_{i}^h|$ in 
Eq.~(\ref{Mixing-coefficients-h-assy}) must be smaller than one. When
$|U_{i+1}^h/U_{i}^h|>1$, the chiral matter disappears at low energy.

\begin{figure}
\centering
\includegraphics[totalheight=3cm]{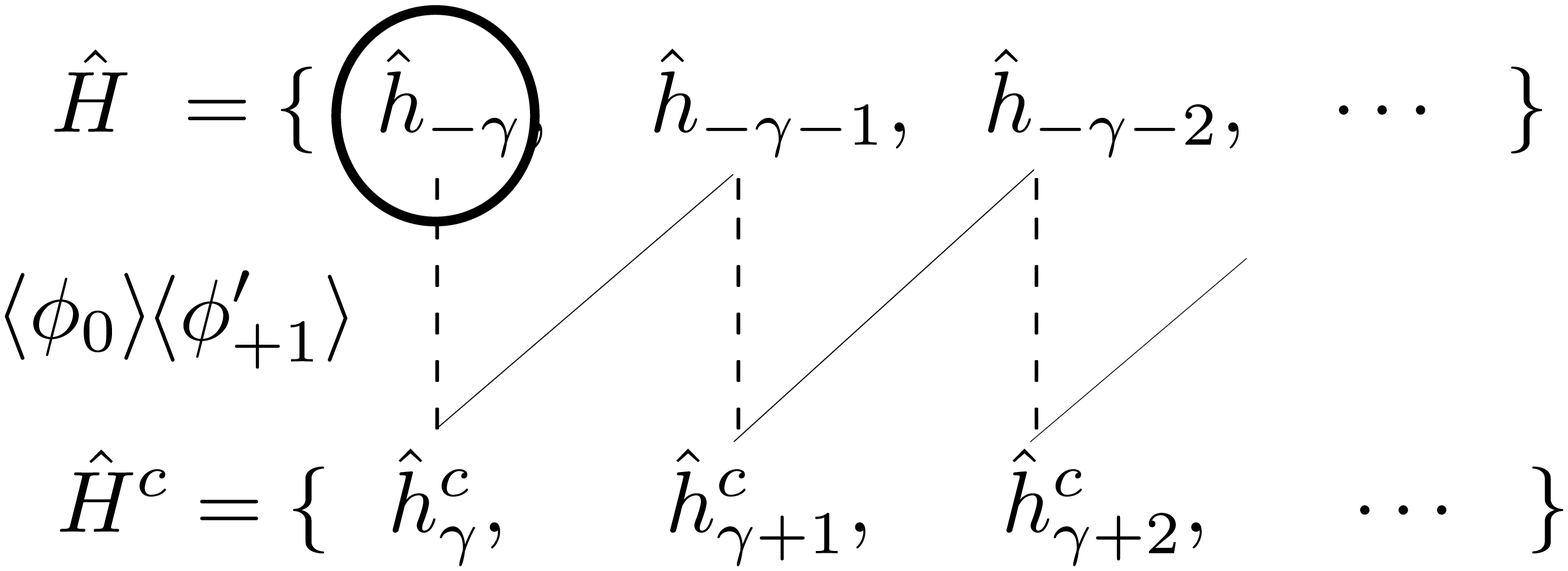}
\includegraphics[totalheight=3cm]{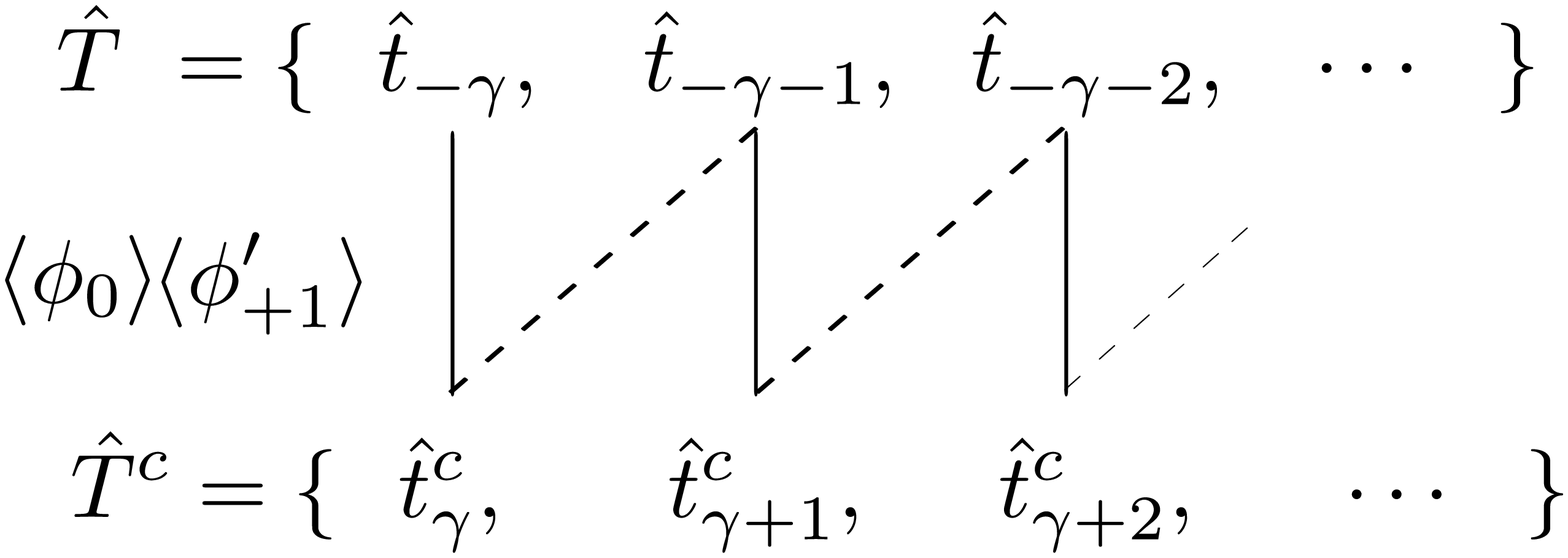}
\caption{The left-hand figure shows that 
when the condition in Eq.~(\ref{doublet-higgs-epsilon}) is satisfied,
there is the one massless
generation of doublet higgses and each massless mode $\hat{h}$ 
realized as certain linear combinations of the components
$\hat{h}_{-\gamma-k}$ 
$(k=0,1,2,\cdots)$, and the constructional element of each massless mode
is determined by its mixing coefficients $U_{k}^h$
in Eq.~(\ref{Mixing-coefficients-h}).
The right-hand figure shows that 
when the condition in Eq.~(\ref{triplet-higgs-epsilon}) is satisfied,
there are no massless generations of
colored higgses. 
The components of the matter fields $\hat{H}$ and $\hat{H}^c$ connected
by lines have a mass term that comes from the VEVs
$\langle\phi_{0}\rangle$ or $\langle\phi_{+1}'\rangle$.
The mass term of the solid line dominantly contributes to whether
massless modes appear or not, and the mass term of the dashed line is
subdominant.
(The mass term of the solid line does not always dominantly contribute
to small components of the mixing coefficients.)
A component surrounded by a circle is a main element of each massless
chiral mode when the mass terms of the solid line dominantly
contribute to small components of the mixing coefficients.}
\label{fig:SGG-min-g=1}
\end{figure}

By using the above normalizable condition, we consider the condition to
realize existence of the up- and down-type doublet higgses and absence
of the up- and down-type colored higgses at low energies.
To produce the up- and down-type higgses at low emeries, 
the parameters $\epsilon_{h_u}$ and $\epsilon_{h_d}$ defined by
\begin{align}
\epsilon_{h_u}:=-2\frac{x_{h_u}\langle\phi_0\rangle}
{z_{h_u}\langle\phi_{+1}'\rangle},\ \ \
\epsilon_{h_d}:=2\frac{x_{h_d}\langle\phi_0\rangle}
{z_{h_d}\langle\phi_{+1}'\rangle}
\end{align}
must satisfy the following conditions
\begin{align}
|\epsilon_{h_u}|,|\epsilon_{h_d}|<
\epsilon_{cr},\ \ \
\epsilon_{cr}:=\sqrt{\frac{S!S!}{(S+1)!(S-1)!}}.
\label{doublet-higgs-epsilon}
\end{align}
To eliminate the up- and down-type colored higgses at low energies, 
the following condition must be satisfied.
\begin{align}
\epsilon_{cr}<|\epsilon_{t_u}|,|\epsilon_{t_d}|,
\end{align}
where the parameters $\epsilon_{t_u}$ and $\epsilon_{t_d}$ are defined
by 
\begin{align}
\epsilon_{t_u}:=3\frac{x_{h_u}\langle\phi_0\rangle}
{z_{h_u}\langle\phi_{+1}'\rangle},\ \ \
\epsilon_{t_d}:=-3\frac{x_{h_d}\langle\phi_0\rangle}
{z_{h_d}\langle\phi_{+1}'\rangle}.
\end{align}
When we rewrite this condition by using $\epsilon_{h_u}$ and
$\epsilon_{h_d}$,  
\begin{align}
\frac{2}{3}\epsilon_{cr}<|\epsilon_{h_u}|,|\epsilon_{h_d}|.
\label{triplet-higgs-epsilon}
\end{align}
Thus, only the up- and down-type higgses appear at low energies if the
parameters $\epsilon_{h_u}$ and $\epsilon_{h_d}$ satisfy the following
condition:
\begin{align}
\frac{2}{3}\epsilon_{cr}<|\epsilon_{h_u}|,|\epsilon_{h_d}|<\epsilon_{cr}.
\end{align}
This is shown in Fig.~\ref{fig:SGG-min-g=1}

\subsection{No chiral generations of others}

\begin{table}[t]
\begin{center}
\begin{tabular}{|c|c|ccc|ccccc|}\hline
Field    
&$\hat{N}$
&$\hat{T}_h$&$\hat{Q}_h$&$\hat{C}_h$
&$\hat{A}_\ell$&$\hat{W}_\ell$&$\hat{G}_\ell$
&$\hat{X}_\ell$&$\hat{Y}_\ell$\\\hline\hline
$SU(3)_C$
&${\bf 1}$
&${\bf 1}$&${\bf 6}$&${\bf 3}$
&${\bf 1}$&${\bf 1}$&${\bf 8}$
&${\bf 3}$&${\bf 3}^*$\\
$SU(2)_L$
&${\bf 1}$
&${\bf 3}$&${\bf 1}$&${\bf 2}$
&${\bf 1}$&${\bf 3}$&${\bf 1}$
&${\bf 2}$&${\bf 2}$\\
$U(1)_Y$
&$0$
&$+1$&$-2/3$&$+1/6$
&$0$&$0$&$0$&$-5/6$&$+5/6$\\\hline
\end{tabular}
\end{center}
\caption{The quantum numbers of 
$G_{SM}=SU(3)_C\times SU(2)_L\times U(1)_Y$ 
for matter fields in the $SU(5)\times SU(1,1)$ model are given in the
table, and their conjugate fields have opposite representations.
$\hat{N}$ belongs to $\hat{N}_{\bf 1}$.
$\hat{T}_h$, $\hat{Q}_h$, $\hat{C}_h$ belong to
$\hat{T}_{\bf 15}$.
$\hat{A}_\ell$, $\hat{W}_\ell$, $\hat{G}_\ell$,
$\hat{X}_\ell$, $\hat{Y}_\ell$ belong to $\hat{A}_{\bf 24}$.
\label{tab:SM-reps:others}}
\end{table}

We consider the $SU(5)$ singlets $\hat{S}_{\bf 1}$ and $\hat{R}_{\bf 1}$
with the positive lowest weights $\eta$ and $\lambda$ of $SU(1,1)$ and
their conjugates 
\begin{align}
W_{\rm M}=&
M_s\hat{S}_{\bf 1}\hat{S}_{\bf 1}^c
+M_r\hat{R}_{\bf 1}\hat{R}_{\bf 1}^c
+x_s\hat{S}_{\bf 1}\hat{S}_{\bf 1}^c\hat{\Phi}_{\bf 1}
+x_r\hat{R}_{\bf 1}\hat{R}_{\bf 1}^c\hat{\Phi}_{\bf 1}
+z_s\hat{R}_{\bf 1}\hat{S}_{\bf 1}^c\hat{\Phi}_{\bf 1}
+z_s'\hat{S}_{\bf 1}\hat{R}_{\bf 1}^c\hat{\Phi}_{\bf 1},
\label{SGG-mu}
\end{align}
where $M$s are mass parameters, $x$s and $z$s are dimensionless coupling
constants. The coupling terms
$\hat{R}_{\bf 1}\hat{S}_{\bf 1}^c\hat{\Phi}_{\bf 1}$ and
$\hat{S}_{\bf 1}\hat{R}_{\bf 1}^c\hat{\Phi}_{\bf 1}$
are allowed if $\Delta_s:=\eta-\lambda$ is an integer and the $SU(1,1)$
spin $S$ of the structure field $\hat{\Phi}_{\bf 1}$ is larger than or
equal to $|\Delta_s|$ ($S\geq|\Delta_s|$).
The nonvanishing VEV $\langle\phi_{0}\rangle$ of the structure field
$\hat{\Phi}_{\bf 1}$ gives additional masses for all components of 
the singlets. Unfortunately, we need fine-tuning between $M_r$ and
$x_r\langle\phi_{0}\rangle$ to generate the first components $\hat{r}$
and $\hat{r}^c$ of the matter fields $\hat{R}_{\bf 1}$ and 
$\hat{R}_{\bf 1}^c$ with the mass $O(m_{\rm SUSY})$.
These light fields are necessary to produce the effective $\mu$-term of
up- and down-type higgses $O(m_{\rm SUSY})$.
Note that the other components of $\hat{R}_{\bf 1}$ and 
$\hat{R}_{\bf 1}^c$ have at least $O(M_{\rm GUT})$ because of the 
difference of the CGC of $SU(1,1)$ between 
$\hat{R}_{\bf 1}\hat{R}_{\bf 1}^c$ and 
$\hat{R}_{\bf 1}\hat{R}_{\bf 1}^c\hat{\Phi}_{\bf 1}$.
In other words, we cannot realize more than one massless vectorlike
matter field at low energy for each matter field.
The other singlets $\hat{S}_{\bf 1}$ and $\hat{S}_{\bf 1}^c$ have the
masses at least $O(M_{\rm GUT})$.

We must emphasize the above fine-tuning problem. This is obviously
unnatural, and this unnaturalness strongly suggests the incompleteness
of this model. To solve the fine-tuning problem, one may prefer to use
tiny mass 
$M_r\sim O(m_{\rm SUSY})$ compared to $O(M_{\rm GUT})$ and tiny
dimensionless coupling constant $x_r\sim O(m_{\rm SUSY}/G_{\rm GUT})$
compared to $O(1)$ without any reason.
Alternatively, models that include additional matter and structure
fields may lead to massless SM singlets $\hat{r}$ and $\hat{r}^c$ via
spontaneous generation of generations without any naturalness problem. 
However, we will not pursue this possibility in this paper. 

We next discuss the fields that are necessary to generate neutrino
masses via seesaw mechanisms. 
First, we consider $SU(5)$ singlet $\hat{N}_{\bf 1}$ with the
positive lowest weight $\xi$ of $SU(1,1)$ and its conjugate.
The superpotential contains 
\begin{align}
W_{\rm N}=&M_n\hat{N}_{\bf 1}\hat{N}^c_{\bf 1}
+x_n\hat{N}_{\bf 1}\hat{N}^c_{\bf 1}\hat{\Phi}_{\bf 1},
\label{SGG-nu-T1}
\end{align}
where $M_n$ is a mass parameter and $x_n$ is a dimensionless coupling
constant.
The same as the fields $\hat{S}_{\bf 1}$ and $\hat{R}_{\bf 1}$, the
nonvanishing VEV $\langle\phi_{0}\rangle$
of the structure field $\hat{\Phi}_{\bf 1}$ 
in Eq.~(\ref{Structure-VEVs}) gives huge masses to all components of
the matter fields $\hat{N}_{\bf 1}$ and  $\hat{N}_{\bf 1}^c$.
Here we assume that the coupling terms such as 
$\hat{N}_{\bf 1}\hat{S}_{\bf 1}\Phi_{\bf 1}$ and 
$\hat{N}_{\bf 1}\hat{R}_{\bf 1}\Phi_{\bf 1}$ 
are forbidden by $R$-parity or the $SU(1,1)$ weight conditions. We will
discuss this in Sec.~\ref{Sec:BLBreaking}.

Second, we consider $SU(5)$ {\bf 15}-plet $\hat{T}_{{\bf 15}}$ with the
negative highest weight $\tau$ of $SU(1,1)$ and its conjugate
because $SU(2)_L$ triplet $\hat{T}_{h}$ is contained in $SU(5)$ 
{\bf 15}-plet $\hat{T}_{{\bf 15}}$ . The superpotential contains  
\begin{align}
W_{\rm N}=&M_t\hat{T}_{\bf 15}\hat{T}^c_{{\bf 15}^*}
+x_t\hat{T}_{\bf 15}\hat{T}^c_{{\bf 15}^*}\hat{\Phi}_{\bf 1}
+z_t\hat{T}_{\bf 15}\hat{T}^c_{{\bf 15}^*}\hat{\Phi}_{\bf 24}',
\label{SGG-nu-T2}
\end{align}
where $M_t$ is a mass parameter, and $x_t$ and $z_t$ are dimensionless
coupling constants. The nonvanishing VEVs $\langle\phi_{0}\rangle$ and
$\langle\phi_{+1}'\rangle$ of the structure fields $\hat{\Phi}_{\bf 1}$
and $\hat{\Phi}_{\bf 24}'$ in Eq.~(\ref{W:higgses:vacuum}) give huge
masses to all components of the matter fields $\hat{T}_{{\bf 15}}$ and
$\hat{T}_{{\bf 15}^*}^c$.  
In this case, the discussion of whether massless particles appear or not
is exactly the same as in the higgs cases. 
The parameter $\epsilon_{t_h}$ defined as
\begin{align}
\epsilon_{t_h}:=\frac{x_{t}\langle\phi_0\rangle}
{z_{t}\langle\phi_{+1}'\rangle}
\end{align}
must satisfy the condition
\begin{align}
|\epsilon_{t_h}|> \epsilon_{cr}
\end{align}
for the triplet higgs $\hat{T}_h$ to disappear at low energy. 
In this case, the other fields $\hat{Q}_h$ and $\hat{C}_h$ in the
$SU(5)$ ${\bf 15}$-plet $\hat{T}_{\bf 15}$ shown in
Table~\ref{tab:SM-reps:others} automatically disappear at low energy
because the triplet higgs has the largest $U(1)_Y$ charge within the
$SU(5)$ ${\bf 15}$-plet. 

One may suspect that, if the coupling terms 
$\hat{F}_{{\bf 10}}\hat{T}_{{\bf 15}^*}^c\hat{\Phi}_{\bf 24}$
and $\hat{F}_{{\bf 10}^*}^c\hat{T}_{\bf 15}\hat{\Phi}_{\bf 24}$
are allowed, they could disturb the structure of the chiral generations
for quarks and leptons. Fortunately, both $\hat{F}_{\bf 10}^{(\prime)}$
and $\hat{T}_{{\bf 15}^*}^c$ belong to positive fields, where positive
fields are chiral superfields with the positive weight of $SU(1,1)$.
Thus these couplings are not allowed.

Third, we consider $SU(5)$ {\bf 24}-plet $\hat{A}_{\bf 24}$ with the
positive lowest weight $\zeta$ of $SU(1,1)$ and its conjugate.
The superpotential contains 
\begin{align}
W_{\rm N}=&M_a\hat{A}_{\bf 24}\hat{A}^c_{\bf 24}
+x_a\hat{A}_{\bf 24}\hat{A}^c_{\bf 24}\hat{\Phi}_{\bf 1}
+z_{sa}\hat{A}_{\bf 24}\hat{A}^c_{\bf 24}\hat{\Phi}_{\bf 24}',
+z_{aa}\hat{A}_{\bf 24}\hat{A}^c_{\bf 24}\hat{\Phi}_{\bf 24}',
\label{SGG-nu-T3}
\end{align}
where $M_a$ is a mass parameter, and $x_a$, $z_{sa}$ and $z_{aa}$ 
are dimensionless coupling constants. 
The last two terms represent the symmetric and anti-symmetric
invariants under $SU(5)$ transformation
built from three fields with the $SU(5)$ adjoint
representation. Note that while the CGCs of the anti-symmetric
invariant are proportional to the $U(1)_Y$ charges, the CGCs of the
symmetric invariant are not proportional to the $U(1)_Y$ charges. 
Also, the CGCs of the invariant built by two adjoint representations are not
proportional to the identity. (See Ref.~\cite{DelOlmo:1987bc} 
for the CGCs of $SU(5)$ adjoint representations in detail.)
Thus, we need to consider the renormalizable condition for the
components of the fields $\hat{A}_\ell$, $\hat{W}_\ell$, $\hat{G}_\ell$,
$\hat{X}_\ell$, $\hat{Y}_\ell$ shown in Table~\ref{tab:SM-reps:others}
and their conjugate fields.
In general, the fields are massive via the nonvanishing VEVs
$\langle\phi_{0}\rangle$ and $\langle\phi_{+1}'\rangle$ of the
structure fields $\hat{\Phi}_{\bf 1}$ and $\hat{\Phi}_{\bf 24}'$ in
Eq.~(\ref{W:higgses:vacuum}) when the parameter $\epsilon_{a}$
defined as   
\begin{align}
\epsilon_{a}:=
\frac{x_{a}\langle\phi_0\rangle}
{(N_iz_{sa}+Y_iz_{aa})\langle\phi_{+1}'\rangle},
\end{align}
satisfies the following condition
\begin{align}
|\epsilon_{a}|> \epsilon_{cr},
\end{align}
where $N_i$ is proportional to a ratio of the CGCs for a basis of
$G_{SM}$ between the singlet built by two adjoint representations and
the SM singlet of the symmetric component built by three adjoint
representations.

\section{Structures of Yukawa couplings}
\label{Sec:Yukawa}

We now discuss the Yukawa couplings between quarks and leptons
and higgses.
\begin{align}
W_Y=y_{\bf 10}\hat{F}_{\bf 10}^{(\prime)}\hat{F}_{\bf 10}^{(\prime)}
\hat{H}_{u{\bf 5}}
+y_{\bf 5}\hat{F}_{\bf 10}^{(\prime)}\hat{G}_{{\bf 5}^*}^{(\prime)}
\hat{H}_{d{\bf 5}^*},
\label{W:Yukawa}
\end{align}
where $y_{\bf 10}$ and $y_{\bf 5}$ are dimensionless coupling
constants. 
Each Yukawa coupling can be classified into two types.
For the first term in Eq.~(\ref{W:Yukawa})
of $\hat{F}_{\bf 10}^{(\prime)}$ and 
$\hat{H}_{u{\bf 5}}$, 
one is $\gamma=2\alpha+[\mbox{positive half-integer}]$;
\begin{align}
W_Y=y_{\bf 10}\hat{F}_{\bf 10}\hat{F}_{\bf 10}'\hat{H}_{u{\bf 5}},
\label{W:Yukawa-10-H}
\end{align}
where $y_{\bf 10}$ is a coupling constant;
the other is $\gamma=2\alpha+[\mbox{semi-positive integer}]$.
\begin{align}
W_Y=y_{\bf 10}\hat{F}_{\bf 10}\hat{F}_{\bf 10}\hat{H}_{u{\bf 5}}
+y_{\bf 10}'\hat{F}_{\bf 10}'\hat{F}_{\bf 10}'\hat{H}_{u{\bf 5}},
\label{W:Yukawa-10-I}
\end{align}
where $y_{\bf 10}'$ is a coupling constant, and the second term is
allowed if $\gamma\geq 2\alpha'$. 
For the second term in Eq.~(\ref{W:Yukawa}), one 
is $\gamma=\alpha+\beta+[\mbox{positive half-integer}]$;
\begin{align}
W_Y=y_{\bf 5}\hat{F}_{\bf 10}'\hat{G}_{{\bf 5}^*}\hat{H}_{d{\bf 5}^*}
+y_{\bf 5}'\hat{F}_{\bf 10}\hat{G}_{{\bf 5}^*}'\hat{H}_{d{\bf 5}^*},
\label{W:Yukawa-5-H}
\end{align}
where $y_{\bf 5}$ and $y_{\bf 5}'$ are coupling constants,
the first term is allowed if $\delta\geq \alpha'+\beta$
and the second term is allowed if $\delta\geq \alpha+\beta'$;
the other is $\delta=\alpha+\beta+[\mbox{semi-positive integer}]$.
\begin{align}
W_Y=y_{\bf 5}\hat{F}_{\bf 10}\hat{G}_{{\bf 5}^*}\hat{H}_{d{\bf 5}^*}
+y_{\bf 5}'\hat{F}_{\bf 10}'\hat{G}_{{\bf 5}^*}'\hat{H}_{d{\bf 5}^*},
\label{W:Yukawa-5-I}
\end{align}
where the second term is allowed if $\delta\geq \alpha'+\beta'$.

We see the structure of the Yukawa couplings in
Eqs.~(\ref{W:Yukawa-10-H})--(\ref{W:Yukawa-5-I}).
After we extract massless modes, we can generally write the
superpotential terms of the Yukawa coupling constant at the low energy 
\begin{align}
W=&y_u^{mn}\hat{q}_m\hat{u}_n^c\hat{h}_u
+y_d^{mn}\hat{q}_m\hat{d}_n^c\hat{h}_d
+y_e^{mn}\hat{\ell}_m\hat{e}_n^c\hat{h}_d.
\end{align}
For the superpotential in Eq.~(\ref{W:Yukawa-10-I}),
the Yukawa coupling constants of up-type quarks are
\begin{align}
&y_u^{mn}=\sum_{i,j=0}^\infty
\left(y_{\bf 10}
C_{i,j}^{\alpha,\alpha,\Delta_\alpha}U_{m,i}^{q}U_{n,j}^u
U_{i+j-\Delta_\alpha}^{h_u}
+y_{\bf 10}'
C_{i,j}^{\alpha',\alpha',\Delta_\alpha'}
U_{m,i}^{q'}U_{n,j}^{u'}U_{i+j-\Delta_\alpha'}^{h_u}
\right),
\end{align}
where $C_{i,j}^{\alpha,\alpha,\Delta_\alpha}$ is a CGC of $SU(1,1)$
given in Ref.~\cite{Yamatsu:2012},
$U$s are given by the spontaneous generation of generations discussed in
Sec.~\ref{Sec:SGG}, $\Delta_\alpha:=\gamma-2\alpha$ and
$\Delta_\alpha':=\gamma-2\alpha'$. 
For the superpotential in Eq.~(\ref{W:Yukawa-10-H}),
the Yukawa coupling constants of up-type quarks are
\begin{align}
&y_u^{mn}=\sum_{i,j=0}^\infty
y_{\bf 10}
\left(C_{i,j}^{\alpha,\alpha',\Delta_\alpha}U_{m,i}^{q}U_{n,j}^{u'}
+C_{i,j}^{\alpha',\alpha,\Delta_\alpha}U_{m,i}^{q'}U_{n,j}^u
\right)
U_{i+j-\Delta_\alpha}^{h_u},
\end{align}
where $\Delta_\alpha:=\gamma-\alpha-\alpha'$.
For the superpotential in Eq.~(\ref{W:Yukawa-5-I}),
the Yukawa coupling constants of down-type quarks and charged leptons are
\begin{align}
&y_d^{mn}=
\sum_{i,j=0}^\infty
\left(
y_{\bf 5}C_{i,j}^{\alpha,\beta,\Delta_\beta}U_{m,i}^{q}U_{n,j}^d
U_{i+j-\Delta_\beta}^{h_d}
+y_{\bf 5}'C_{i,j}^{\alpha',\beta',\Delta_\beta'}U_{m,i}^{q'}U_{n,j}^{d'}
U_{i+j-\Delta_\beta'}^{h_d}
\right),\\
&y_e^{mn}=
\sum_{i,j=0}^\infty
\left(
y_{\bf 5}C_{i,j}^{\beta,\alpha,\Delta_\beta}
U_{m,i}^{\ell}U_{n,j}^{e}U_{i+j-\Delta_\beta}^{h_d}
+y_{\bf 5}'C_{i,j}^{\beta',\alpha',\Delta_\beta'}
U_{m,i}^{\ell'}U_{n,j}^{e'}U_{i+j-\Delta_\beta'}^{h_d}
\right),
\end{align}
where $\Delta_\beta:=\delta-\alpha-\beta$ and
$\Delta_\beta':=\delta-\alpha'-\beta'$.
For the superpotential in Eq.~(\ref{W:Yukawa-5-H}),
the Yukawa coupling constants of down-type quarks and charged leptons are
\begin{align}
&y_d^{mn}=
\sum_{i,j=0}^\infty
\left(
y_{\bf 5}C_{i,j}^{\alpha',\beta,\Delta_\beta}U_{m,i}^{q'}U_{n,j}^d
U_{i+j-\Delta_\beta}^{h_d}
+y_{\bf 5}'C_{i,j}^{\alpha,\beta',\Delta_\beta'}U_{m,i}^{q}U_{n,j}^{d'}
U_{i+j-\Delta_\beta'}^{h_d}
\right),\\
&y_e^{mn}=
\sum_{i,j=0}^\infty
\left(
y_{\bf 5}
C_{i,j}^{\beta,\alpha',\Delta_\beta}
U_{m,i}^{\ell}U_{n,j}^{e'}U_{i+j-\Delta_\beta}^{h_d}
+y_{\bf 5}'C_{i,j}^{\beta',\alpha,\Delta_\beta'}
U_{m,i}^{\ell'}U_{n,j}^{e}U_{i+j-\Delta_\beta'}^{h_d}
\right),
\end{align}
where $\Delta_\beta:=\delta-\alpha'-\beta$ and
$\Delta_\beta':=\delta-\alpha-\beta'$.

The Yukawa coupling constants are completely determined by the overall
couplings $y$s and the mixing coefficients $U$s.
In particular, the weight condition satisfies $\gamma=\alpha+\alpha'$,
$\delta=\alpha'+\beta$, and  $q_\alpha<q_\beta$.
Each Yukawa coupling matrix has only one overall coupling constant.

We consider the mixing coefficients of down-type quarks and charged
leptons given in Eqs.~(\ref{MC:quarks-leptons-1}) and
(\ref{MC:quarks-leptons-2}). For nonzero coupling constants $z$ and $w$,
the mixing coefficients are different because 
the $U(1)_Y$ charges of down-type quarks are different from those of
charged leptons. 
Thus, the Yukawa coupling constants of down-type quarks can be
different from those of charged leptons.

The patterns of the mixing coefficients are highly dependent on the
values of $q_\alpha$ and $q_\beta$ that determine dominant massless
components. A detailed investigation of the Yukawa couplings is not
the purpose in this paper, so we will not analyze the mass eigenvalues
of quarks and leptons, and the CKM
\cite{Cabibbo:1963yz,Kobayashi:1973fv} and MNS \cite{Maki:1962mu} 
matrices. One can find the basic argument in
Refs.~\cite{Inoue:1994qz,Inoue:2000ia,Yamatsu:2007,Yamatsu:2012}.

\section{$\mu$-term}
\label{Sec:mu-term}

We need to generate the effective $\mu$-term $\mu\hat{h}_u\hat{h}_d$,
where $\mu\simeq O(m_{\rm SUSY})$ is the supersymmetry breaking mass
parameter $O(10^{2\sim 3})$ GeV \cite{Yamatsu:2008}.
This is because the $\mu$-term $\mu\hat{H}_u\hat{H}_d$ is forbidden by
the noncompact horizontal symmetry $G_N=SU(1,1)$ since both chiral
higgses $\hat{h}_u$ and $\hat{h}_d$ 
are contained in the negative fields $\hat{H}_u$ and $\hat{H}_d$
\cite{Inoue:1994qz,Inoue:1996te}, 
where negative fields are chiral superfields with the negative weight
of $SU(1,1)$.
To generate the effective $\mu$-term, the up- and down-type higgses
$\hat{H}_u$ and $\hat{H}_d$ must couple to a positive field $\hat{S}$
belonging to the singlet under $G_{\rm SM}$, and the field must get a
nonvanishing VEV $O(m_{\rm SUSY})$. Unlike the Next-to-minimal
supersymmetric SM (NMSSM) that contains an
extra singlet superfield under $G_{\rm SM}$, the horizontal symmetry
does not allow the existence of linear, quadratic and cubic terms, e.g.,
$M^2\hat{S}$, $M\hat{S}^2$, and $\lambda\hat{S}^3$. Thus, in this model,
we cannot use the same method as in the NMSSM.

If the up- and down-type higgses $\hat{h}_u$ and $\hat{h}_d$ belong to
conjugate representations or the same real representation,
then the effective  $\mu$-term $\mu\hat{h}_u\hat{h}_d$ is generated
only by singlet fields and the VEVs of 
the intermediate scale $O(\sqrt{m_{\rm SUSY}M_{\rm GUT}})$ between the
supersymmetry 
breaking mass scale $O(m_{\rm SUSY})$ and the fundamental scale
$O(M_{\rm GUT})$. 
If the up- and down-type higgses do not belong to conjugate
representations, then the effective $\mu$-term needs to be
generated not only by singlet representations of compact unified group
but also non-singlet representations. The seesaw mechanism
between the fundamental scale $O(M_{\rm GUT})$ and the intermediate scale
$O(\sqrt{m_{\rm SUSY}M_{\rm GUT}})$ generates the supersymmetry breaking
mass scale $O(m_{\rm SUSY})$. 

Let us first consider how to generate the non-vanishing VEVs with the
supersymmetry breaking mass scale $O(m_{\rm SUSY})$ and the intermediate
scale $O(\sqrt{m_{\rm SUSY}M_{\rm GUT}})$ from the fundamental scale
$O(M_{\rm GUT})$ and  the supersymmetry breaking mass scale $O(m_{\rm
SUSY})$. 
To realize this situation, we need to introduce some matter fields that
are singlets under $SU(5)$. 

The simplest superpotential contains the $SU(5)$
singlets $\hat{S}_{\bf 1}$ and $\hat{R}_{\bf 1}$ with the positive
lowest weights $\eta=\gamma+\delta$ and $\lambda=(\gamma+\delta)/2$ of 
$SU(1,1)$ and their conjugates 
\begin{align}
W_{\rm M}=
y\hat{R}_{\bf 1}\hat{R}_{\bf 1}\hat{S}_{\bf 1}^c
+y^c\hat{R}_{\bf 1}^c\hat{R}_{\bf 1}^c\hat{S}_{\bf 1}
+y'\hat{H}_{u{\bf 5}}\hat{H}_{d{\bf 5}^*}\hat{S}_{\bf 1}
+y^{\prime c}\hat{H}_{u{\bf 5}^*}^c\hat{H}_{d{\bf 5}}^c\hat{S}_{\bf 1}^c,
\end{align}
where $y$s are coupling constants.
From the above superpotential and the superpotential in
Eq.~(\ref{SGG-mu}), decoupling the singlets except 
the first component of the singlets 
$\hat{S}_{\bf 1}$, $\hat{S}_{\bf 1}^c$, $\hat{R}_{\bf 1}$ 
and $\hat{R}_{\bf 1}^c$, we obtain 
\begin{align}
W_{\rm M}=
\tilde{M}_s\hat{s}\hat{s}^c
+\tilde{M}_r\hat{r}\hat{r}^c
+y\hat{r}\hat{r}\hat{s}^c
+y^c\hat{r}^c\hat{r}^c\hat{s}
+y'U_0^{h_u}U_0^{h_d}\hat{h}_{u{\bf 5}}\hat{h}_{d{\bf 5}^*}\hat{s},
\end{align}
where we assume 
$\tilde{M}_s:=M_s+x_s\langle\phi_0\rangle\sim O(M_{\rm GUT})$ and 
$\tilde{M}_r:=M_r+x_r\langle\phi_0\rangle\sim O(m_{\rm SUSY})$,
the $U_{0}^{h_u}$ and $U_{0}^{h_d}$ are mixing coefficients of the
up- and down-type higgses.
Decoupling $\hat{s}$ and $\hat{s}^c$ by using 
\begin{align}
\frac{\partial W}{\partial\hat{s}}=
\tilde{M}_s\hat{s}^c+y^c\hat{r}^c\hat{r}^c
+y'U_0^{h_u}U_0^{h_d}\hat{h}_{u{\bf 5}}\hat{h}_{d{\bf 5}^*}=0,\ \ \
\frac{\partial W}{\partial\hat{s}^c}=
\tilde{M}_s\hat{s}+y\hat{r}\hat{r}=0,
\end{align}
we have
\begin{align}
W_{\rm M}=
\tilde{M}_r\hat{r}\hat{r}^c
-\frac{y}{\tilde{M}_s}\hat{r}\hat{r}
\left(y^c\hat{r}^c\hat{r}^c
+y'U_0^{h_u}U_0^{h_d}\hat{h}_{u{\bf 5}}\hat{h}_{d{\bf 5}^*}\right).
\end{align}
This leads to the scalar potential 
\begin{align}
V_{\rm SUSY}=&\left|\frac{\partial W}{\partial r}\right|^2
+\left|\frac{\partial W}{\partial r^c}\right|^2
=\left|\tilde{M}_rr^c-2\frac{yy^c}{\tilde{M}_s}rr^cr^c\right|^2
+\left|\tilde{M}_rr-2\frac{yy^c}{\tilde{M}_s}rrr^c\right|^2.
\end{align}
Its corresponding SUSY breaking terms are 
\begin{align}
V_{\cancel{\rm SUSY}}=
B_r\tilde{M}_rrr^c-A_r\frac{yy^c}{\tilde{M}_s}rrr^cr^c+\mbox{h.c.}
+\tilde{m}_r^2|r|^2+\tilde{m}_r^{c2}|r^c|^2,
\end{align}
where $B_r$ is a $B$-parameter of $\hat{r}$ and $\hat{r}^c$,  $A_r$ is
an $A$-parameter of $\hat{r}\hat{r}\hat{s}$, and $\tilde{m}_r^2$ 
and $\tilde{m}_r^{c2}$ 
are soft masses of $\hat{r}$ and $\hat{r}^c$, respectively.
The total scalar potential is 
\begin{align}
V=&V_{\rm SUSY}+V_{\cancel{\rm SUSY}}.
\end{align}

After we perform tedious calculation, we obtain 
$\langle r\rangle, \langle r^c\rangle=O(\sqrt{m_{\rm SUSY}M_{\rm GUT}})$
and $\langle s\rangle=O(m_{\rm SUSY})$ as discussed in
Ref.~\cite{Yamatsu:2008}.
Thus, the effective $\mu$-term between $h_{u{\bf 5}}$ and $h_{d{\bf 5}}$ 
is $O(m_{\rm SUSY})$.
The singlet fermions and scalars $\hat{r}$ and $\hat{r}^c$ 
have a mass term $O(m_{\rm SUSY})$ except the Nambu-Goldstone (NG) boson
since this potential have a $U(1)$ global symmetry at low energy and
this symmetry is broken by the nonvanishing VEVs of the singlets.
Note that, if there is no SUSY breaking term, the singlet fermion is
massless because SUSY forces the fermionic partner of the NG boson to
be a pseudo-NG fermion
\cite{Bando:1984cc,Bando:1984fn,Inoue:1985cw,Bando2003}. 

In addition, the coupling between the higgses $\hat{h}_{u{\bf 5}}$ and
$\hat{h}_{d{\bf 5}^*}$ and the singlets is suppressed by the
factor $O(\sqrt{m_{\rm SUSY}/M_{\rm GUT}})$. 
Therefore, the effective theory below the energy scale 
$\sqrt{m_{\rm SUSY}M_{\rm GUT}}$ is described by the MSSM and the almost
decoupled $G_{\rm SM}$ singlets.

The NG boson may cause some problems for cosmology, e.g., a moduli
problem \cite{Lyth:1995ka}.  To solve the moduli problem, we should
assume that there is thermal inflation after reheating takes place as
discussed in Ref.~\cite{Lyth:1995ka}. We will not discuss the
cosmological problems in this paper.

\section{Baryon and/or lepton number violating terms}
\label{Sec:BLBreaking}

We classify the baryon and/or lepton number violating terms 
up to superpotential quartic order by using $SU(1,1)$ symmetry and the
$R$-parity \cite{Fayet:1977yc}  (matter parity \cite{Bento:1987mu})
shown in Table~\ref{tab:matter}. For a review, see, e.g.,
Ref.~\cite{Barbier:2004ez}. In the following, we omit the mirror terms.
$\lambda$s stand for dimensionless couplings,  $\Lambda$s and 
$\mu$ are dimension-one parameters, $\Delta$s are integer, and
$\Delta_\pm$ is a non-negative integer.

To make the invariants under the $SU(1,1)$ transformation, 
we can use the following way; first, we make the composite states of
only positive field or negative field. 
In general, a composite field built by multi-positive fields
$\hat{F}_i$ $(i=0,1,2,\cdots)$ with the lowest weight $\alpha_i$ is a
positive field with the lowest weight $\sum_i\alpha_i+\Delta_+$
($\Delta_+=0,1,2,\cdots$). 
A composite field built by multi-negative fields
$\hat{H}_j$ $(j=0,1,2,\cdots)$ with the highest weight $-\beta_j$ is a
negative field with the highest weight $-\sum_j\beta_j-\Delta_-$
($\Delta_-=0,1,2,\cdots$). 
When the multi-positive field contains only one positive field,
$\Delta_+=0$;
when the multi-negative field contains only one negative field,
$\Delta_-=0$.
Next, we combine the multi-positive and negative fields.
The invariants built by the multi-positive and negative fields must
satisfy the condition 
$\sum_i\alpha_i+\Delta_+=\sum_j\beta_j+\Delta_-$:
i.e., $\Delta:=\Delta_+-\Delta_-=\sum_j\beta_j-\sum_i\alpha_i$.
We define $\Delta$ as the difference between the sum of the lowest
weights of positive fields and the highest weights of negative fields,
where a positive field is a matter field with only the positive
weights of $SU(1,1)$ and a negative field is a matter field with only
the negative weights of $SU(1,1)$.
More explicitly, 
for a term containing one positive field with the lowest
weight $\alpha$ and one negative field with the highest weight
$-\beta$, the condition $\alpha=\beta$ must be satisfied;
for a term containing two positive fields with the lowest
weights $\alpha$ and $\alpha'$ and one negative field with the
highest weight $-\beta$, the condition 
$\Delta=\Delta_+=\beta-\alpha-\alpha'$ must be satisfied;
for a term containing three positive fields with the
lowest weights $\alpha$, $\alpha'$ and $\alpha''$ and one negative field
with the highest weight $-\beta$, the condition 
$\Delta=\Delta_+=\beta-\alpha-\alpha'-\alpha''$
must be satisfied;
for a term containing two positive fields with the lowest
weights $\alpha$ and $\alpha'$ and two negative fields with the
highest weights $-\beta$ and $-\beta'$, the condition 
$\Delta=\Delta_+-\Delta_-=\beta+\beta'-\alpha-\alpha'$ must be satisfied. 

We start to consider the $SU(5)$ GUT model with $SU(1,1)$.
First, $SU(1,1)$ symmetry and $R$-parity allow the following \cancel{B}
and/or \cancel{L} quartic term 
\begin{align}
W_{M:4;\cancel{B},\cancel{L}}=&
\frac{1}{\Lambda}
\hat{G}_{{\bf 5}^*}^{(\prime)}
\hat{G}_{{\bf 5}^*}^{(\prime)}
\hat{H}_{u{\bf 5}}\hat{H}_{u{\bf 5}}
=\frac{1}{\Lambda}
\hat{L}^{(\prime)}\hat{L}^{(\prime)}\hat{H}_{u}\hat{H}_{u}
+\frac{1}{\Lambda}
\hat{D}^{c(\prime)}\hat{D}^{c(\prime)}\hat{T}_{u}\hat{T}_{u}
+\frac{1}{\Lambda}
\hat{D}^{c(\prime)}\hat{L}^{(\prime)}\hat{T}_{u}\hat{H}_{u}
\label{W:Neutrino-mass}
\end{align}
if the $SU(1,1)$ weights satisfy a condition 
$\Delta=\Delta_+-\Delta_-=2\gamma-\beta^{(\prime)}-\beta^{(\prime)}$.
More explicitly, when $\Delta=\Delta_+-\Delta_-=2\gamma-2\beta$, 
$\hat{G}_{{\bf 5}^*}\hat{G}_{{\bf 5}^*}\hat{H}_{u{\bf 5}}\hat{H}_{u{\bf 5}}$ 
are allowed; when $\Delta=\Delta_+-\Delta_-=2\gamma-\beta-\beta^{\prime}$,
$\hat{G}_{{\bf 5}^*}\hat{G}_{{\bf 5}^*}^{\prime}\hat{H}_{u{\bf 5}}\hat{H}_{u{\bf 5}}$ are allowed;
when $\Delta=\Delta_+-\Delta_-=2\gamma-2\beta^{\prime}$, 
$\hat{G}_{{\bf 5}^*}^{\prime}\hat{G}_{{\bf 5}^*}^{\prime}\hat{H}_{u{\bf 5}}\hat{H}_{u{\bf 5}}$
are allowed. In the following we also use the same rule.
Second, $SU(1,1)$ symmetry prohibits and $R$-parity allows the following
\cancel{B} and/or \cancel{L} quartic term 
\begin{align}
W_{M:4;\cancel{B},\cancel{L}}=
\frac{1}{\Lambda}
\hat{F}_{{\bf 10}}^{(\prime)}\hat{F}_{{\bf 10}}^{(\prime)}
\hat{F}_{{\bf 10}}^{(\prime)}\hat{G}_{{\bf 5}^*}^{(\prime)}
= \frac{1}{\Lambda}\hat{Q}^{(\prime)}\hat{Q}^{(\prime)}
\hat{Q}^{(\prime)}\hat{L}^{(\prime)}
+\frac{1}{\Lambda}\hat{U}^{(\prime)c}\hat{U}^{(\prime)c}
\hat{D}^{(\prime)c}\hat{E}^{(\prime)c}
\label{W:Proton-decay}
\end{align}
because $\hat{F}_{{\bf 10}}^{(\prime)}$ and 
$\hat{G}_{{\bf 5}^*}^{(\prime)}$ belong to positive fields.
Third,  $SU(1,1)$ symmetry and $R$-parity prohibit the following
\cancel{B} and/or \cancel{L} cubic term 
\begin{align}
W_{M:3;\cancel{B},\cancel{L}}=&
\lambda\hat{F}_{{\bf 10}}^{(\prime)}
\hat{G}_{{\bf 5}^*}^{(\prime)}\hat{G}_{{\bf 5}^*}^{(\prime)}
=\lambda\hat{E}^{(\prime)c}\hat{L}^{(\prime)}\hat{L}^{(\prime)}
+\lambda\hat{Q}^{(\prime)}\hat{D}^{(\prime)c}\hat{L}^{(\prime)}
+\lambda\hat{U}^{(\prime)c}\hat{D}^{(\prime)c}\hat{D}^{(\prime)c}
\end{align}
because $\hat{F}_{{\bf 10}}^{(\prime)}$ and 
$\hat{G}_{{\bf 5}^*}^{(\prime)}$ belong to positive fields.
Finally,  $SU(1,1)$ symmetry allows and $R$-parity prohibits the
following \cancel{B} and/or \cancel{L} quadratic term 
\begin{align}
W_{M:2\cancel{B},\cancel{L}}=&
\mu\hat{G}_{{\bf 5}^*}^{(\prime)}\hat{H}_{u{\bf 5}}
=\mu\hat{L}^{(\prime)}\hat{H}_{u}+\mu\hat{D}^{(\prime)c}\hat{T}_{u}
\end{align}
if $\beta^{(\prime)}=\gamma$. 
The cubic terms are 
\begin{align}
W_{M:3;\cancel{B},\cancel{L}}=&
\lambda\hat{G}_{{\bf 5}^*}^{(\prime)}\hat{H}_{u{\bf 5}}\hat{S}_{\bf 1}
+\lambda'\hat{G}_{{\bf 5}^*}^{(\prime)}\hat{H}_{u{\bf 5}}\hat{S}_{\bf 1}^c
+\lambda''\hat{G}_{{\bf 5}^*}^{(\prime)}\hat{H}_{u{\bf 5}}\hat{R}_{\bf 1}
+\lambda'''\hat{G}_{{\bf 5}^*}^{(\prime)}\hat{H}_{u{\bf 5}}\hat{R}_{\bf 1}^c
\nonumber\\
=&
\lambda\hat{L}^{(\prime)}\hat{H}_u\hat{S}
+\lambda\hat{D}^{(\prime)}\hat{T}_u\hat{S}
+\lambda'\hat{L}^{(\prime)}\hat{H}_u\hat{S}^c
+\lambda'\hat{D}^{(\prime)}\hat{T}_u\hat{S}^c
\nonumber\\
&+\lambda''\hat{L}^{(\prime)}\hat{H}_u\hat{R}
+\lambda''\hat{D}^{(\prime)}\hat{T}_u\hat{R}
+\lambda'''\hat{L}^{(\prime)}\hat{H}_u\hat{R}^c
+\lambda'''\hat{D}^{(\prime)}\hat{T}_u\hat{R}^c,
\label{W:D3:SR}
\end{align}
if $\gamma=\eta+\beta^{(\prime)}+\Delta_+$,
$\beta^{(\prime)}=\eta+\gamma+\Delta_-$,
$\gamma=\lambda+\beta^{(\prime)}+\Delta_+$, and
$\beta^{(\prime)}=\lambda+\gamma+\Delta_-$;
\begin{align}
W_{M:3;\cancel{B},\cancel{L}}=&
\lambda\hat{F}_{{\bf 10}}^{(\prime)}
\hat{H}_{d{\bf 5}^*}\hat{H}_{d{\bf 5}^*}
+\lambda'\hat{F}_{\bf 10}^{(\prime)}\hat{F}_{\bf 10}^{(\prime)}
\hat{G}_{{\bf 5}}^{(\prime)c}\nonumber\\
=&\lambda\hat{E}^{(\prime)c}\hat{H}_d\hat{H}_d
+\lambda\hat{Q}^{(\prime)}\hat{T}_d\hat{H}_d
+\lambda\hat{U}^{(\prime)c}\hat{T}_d\hat{T}_d
+\lambda'\hat{Q}^{(\prime)}\hat{Q}^{(\prime)}\hat{D}^{(\prime)}
+\lambda'\hat{Q}^{(\prime)}\hat{U}^{(\prime)c}\hat{L}^{(\prime)c},
\end{align}
if $\alpha^{(\prime)}=2\gamma+\Delta_-$ and 
$\beta^{(\prime)}=\alpha^{(\prime)}+\alpha^{(\prime)}+\Delta_+$.
The quartic terms are 
\begin{align}
W_{M:4;\cancel{B},\cancel{L}}=&
\frac{1}{\Lambda}
\hat{F}_{{\bf 10}}^{(\prime)}\hat{G}_{{\bf 5}^*}^{(\prime)}
\hat{G}_{{\bf 5}^*}^{(\prime)}\hat{S}_{{\bf 1}}^c
+\frac{1}{\Lambda'}
\hat{F}_{{\bf 10}}^{(\prime)}\hat{G}_{{\bf 5}^*}^{(\prime)}
\hat{G}_{{\bf 5}^*}^{(\prime)}\hat{R}_{{\bf 1}}^c
+\frac{1}{\Lambda''}
\hat{F}_{{\bf 10}}^{(\prime)}\hat{F}_{{\bf 10}}^{(\prime)}
\hat{F}_{{\bf 10}}^{(\prime)}\hat{H}_{d{\bf 5}^*}
+\frac{1}{\Lambda'''}\hat{G}_{{\bf 5}^*}^{(\prime)}\hat{H}_{d{\bf 5}^*}
\hat{H}_{u{\bf 5}}\hat{H}_{u{\bf 5}}\nonumber\\
=&
\frac{1}{\Lambda}
\hat{E}^{(\prime)c}\hat{L}^{(\prime)}\hat{L}^{(\prime)}\hat{S}^c
+\frac{1}{\Lambda}
\hat{Q}^{(\prime)}\hat{D}^{(\prime)c}\hat{L}^{(\prime)}\hat{S}^c
+\frac{1}{\Lambda}
\hat{U}^{(\prime)c}\hat{D}^{(\prime)c}\hat{D}^{(\prime)c}\hat{S}^c
\nonumber\\
&+\frac{1}{\Lambda'}
\hat{E}^{(\prime)c}\hat{L}^{(\prime)}\hat{L}^{(\prime)}\hat{R}^c
+\frac{1}{\Lambda'}
\hat{Q}^{(\prime)}\hat{D}^{(\prime)c}\hat{L}^{(\prime)}\hat{R}^c
+\frac{1}{\Lambda'}
\hat{U}^{(\prime)c}\hat{D}^{(\prime)c}\hat{D}^{(\prime)c}\hat{R}^c
\nonumber\\
&+\frac{1}{\Lambda''}
\hat{Q}^{(\prime)}\hat{Q}^{(\prime)}\hat{Q}^{(\prime)}\hat{H}_d
+\frac{1}{\Lambda''}
\hat{Q}^{(\prime)}\hat{U}^{(\prime)c}\hat{E}^{(\prime)c}\hat{H}_d
\nonumber\\
&+\frac{1}{\Lambda'''}\hat{L}^{(\prime)}\hat{H}_d\hat{H}_u\hat{H}_u
+\frac{1}{\Lambda'''}\hat{D}^{c(\prime)}\hat{T}_d\hat{T}_u\hat{T}_u
+\frac{1}{\Lambda'''}\hat{L}^{(\prime)}\hat{T}_d\hat{H}_u\hat{T}_u
\end{align}
if 
$\lambda=\alpha^{(\prime)}+\alpha^{(\prime)}+\beta^{(\prime)}+\Delta_+$,
$\eta=\alpha^{(\prime)}+\alpha^{(\prime)}+\beta^{(\prime)}+\Delta_+$,
$\delta=\alpha^{(\prime)}+\alpha^{(\prime)}+\alpha^{(\prime)}+\Delta_+$,
and $\beta^{(\prime)}=2\gamma+\delta+\Delta_-$.

In general, SUSY models with $R$-parity violating terms suffer from
rapid proton decay and lepton flavor violations \cite{Barbier:2004ez}. 
Thus, to prevent the unacceptable predictions, the $R$-parity must be
realized at low energy.
Fortunately, even when we discuss SUSY models with $R$-parity
that contain the relevant or marginal terms, after some heavy particles
are integrated out, the effective neutrino ``mass'' term in
Eq.~(\ref{W:Neutrino-mass}) can be induced. Unfortunately, the problematic
operator in Eq.~(\ref{W:Proton-decay}) can be also induced. 

On the other hand, the $SU(1,1)$ horizontal symmetry does not allow the
problematic term in Eq.~(\ref{W:Proton-decay}). Of course, 
once the symmetry is broken, there is no reason to deny generating the
term. We will discuss this topic in this section.

Another interesting feature is that special weight assignments of
$SU(1,1)$ mean that $R$-parity remains even after the $SU(1,1)$ symmetry
is broken. 
One assignment is the following:
\begin{align}
&\alpha=\frac{2n+1}{4},\ 
\alpha'=\frac{2n+1}{4}+q_\alpha,\ 
\beta=\frac{2n+1}{4}+2m,\ 
\beta'=\frac{2n+1}{4}+2m+q_\beta,
\nonumber\\
&\gamma=n+q_\alpha+\frac{1}{2},\ 
\delta=n+2m+q_\alpha+\frac{1}{2},\ 
\eta/2=\lambda=n+m+q_\alpha+\frac{1}{2},
\label{Assignment:SU(1,1)}
\end{align}
where the $SU(1,1)$ weight, such as $\alpha$, must be a positive number, 
$n$ and $m$ are integer, $q_\alpha$ and $q_\beta$ are half-integer.
In other words, the quark and lepton superfields have the quarter values 
of the $SU(1,1)$ weight, and the higgs and the other superfields have
integer values of the $SU(1,1)$ weight. Thus, even numbers of quarks and
leptons are necessary to couple higgses and the other fields. This is
completely the same as the $R$-parity shown in Table~\ref{tab:matter}.
When we construct models with an $SU(1,1)$ horizontal symmetry,
we do not always assume the $R$-parity to prevent rapid proton decay,
lepton flavor violation and to make dark matter candidate.
Note that the assignment is compatible with the Yukawa
couplings in Eqs.~(\ref{W:Yukawa-10-H}) and (\ref{W:Yukawa-5-H}),
but incompatible with those in Eqs.~(\ref{W:Yukawa-10-I}) and
(\ref{W:Yukawa-5-I}).

Also, another example is the following assignment
\begin{align}
&\alpha=\frac{2n+1}{4},\ 
\alpha'=\frac{2n+1}{4}+q_\alpha,\ 
\beta=\frac{2n-3}{4}+2m-3q_\alpha-q_\beta,\ 
\beta'=\frac{2n-3}{4}+2m-3q_\alpha,
\nonumber\\
&\gamma=n+2q_\alpha+\frac{1}{2},\ 
\delta=n+2m-2q_\alpha-\frac{1}{2},\ 
\eta/2=\lambda=n+m,
\label{Assignment:SU(1,1)-2}
\end{align}
where $n$, $m$ are integer.
This assignment is compatible with those in Eqs.~(\ref{W:Yukawa-10-I})
and (\ref{W:Yukawa-5-I}), but incompatible with the Yukawa couplings in
Eqs.~(\ref{W:Yukawa-10-H}) and (\ref{W:Yukawa-5-H}).
The same as the assignment in Eq.~(\ref{Assignment:SU(1,1)}),
this assignment forbids all $R$-parity violating terms because 
the quarks and leptons have the quarter values of the $SU(1,1)$ weight,
the higgses have the half-integer values, and the singlets have the
integer values.

We can find other assignments of $SU(1,1)$ weights to prohibit $R$-parity
violating terms, and to allow the ``neutrino mass'' term in
Eq.~(\ref{W:Neutrino-mass}). The above two assignments in
Eq.~(\ref{Assignment:SU(1,1)}) and Eq.~(\ref{Assignment:SU(1,1)-2})
include enough assignments for the following discussion. We only consider
the model with these $SU(1,1)$ assignment or explicitly imposed 
$R$-parity shown in Table~\ref{tab:matter}. We focus on the
superpotential terms in Eqs.~(\ref{W:Neutrino-mass}) and
(\ref{W:Proton-decay}). 
We will discuss how to obtain sizable neutrino masses and how to
suppress rapid proton decay in this model.

\subsection{Neutrino masses}

We now discuss seesaw mechanisms, so-called Type-I
\cite{Yanagida:1979as},
Type-II \cite{Magg:1980ut,Lazarides:1980nt,Mohapatra:1980yp},
and Type-III \cite{Foot:1988aq} seesaw mechanisms in the MSSM plus
additional necessary field content.
Type-I, Type-II, and Type-III seesaw mechanisms can be achieved
by using right-handed neutrinos $\hat{N}$ with $({\bf 1},{\bf 1},0)$ 
under $G_{SM}$, 
charged triplet higgses $\hat{T}_h$ and $\hat{T}_h^c$ 
with $({\bf 1},{\bf 3},\pm 1)$ under $G_{SM}$, 
and neutral triplet leptons $\hat{W}_\ell$ with $({\bf 1},{\bf 3},0)$, 
where the first, second and third columns stand for an $SU(3)_C$
weight, an $SU(2)_L$ weight, and a $U(1)_Y$ charge, respectively.
We can also classify Type-I and Type-III seesaw mechanisms as
Majorana-type seesaw mechanisms and Type-II as non-Majorana-type.
For a review, see, e.g., Ref.~\cite{Mohapatra:2005wg}.

Each additional field has the following superpotential terms,
respectively
\begin{align}
W_{I}=&\sum_{a,b}M_{N}^{ab}\hat{N}_a\hat{N}_b
+\sum_{i,a}y_I^{ia}\hat{L}^i\hat{H}_u\hat{N}_a,
\label{W:neutrino-Type-I}\\
W_{II}=&\sum_{a,b}M_{T_h}^{ab}\hat{T}_{ha}\hat{T}_{hb}^c
+\sum_{i,j,a}y_{II}^{ija}\hat{L}_i\hat{L}_j\hat{T}_{ha}
+\sum_{a}y_{II}^{\prime a}\hat{H}_u\hat{H}_u\hat{T}_{ha},
\label{W:neutrino-Type-II}\\
W_{III}=&\sum_{a,b}M_{W_\ell}^{ab}\hat{W}_{\ell a}\hat{W}_{\ell b} 
+\sum_{i,a}y_{III}^{ia}\hat{L}_i\hat{H}_u\hat{W}_{\ell a},
\label{W:neutrino-Type-III}
\end{align}
where $M$s are mass parameters, $y$s are coupling constant, $a,b$ stand
for the label of the additional matter fields, and $i$ is the label of 
the left-handed neutrino.
If we assume $M_X$ is much larger than electro-weak scale, after
decoupling the additional fields, we obtain the effective neutrino-higgs
superpotential term 
\begin{align}
W_{\rm eff}=\sum_{i,j=1}^3\frac{\kappa^{ij}}{M_X}
\hat{L}_i\hat{L}_j\hat{H}_u\hat{H}_u,
\end{align}
where $M_X$ is a mass parameter and $\kappa^{ij}$ is a coupling constant
matrix determined by the mass parameters and the coupling constants 
in Eq.~(\ref{W:neutrino-Type-I})--(\ref{W:neutrino-Type-III}).
After the up-type higgs $\hat{H}_u$ obtains a non-vanishing VEV, the
coupling term becomes the mass term of the left-handed neutrinos.
If $M_X$ is $O(M_{\rm GUT})$, the effective masses become 
$O(v_{\rm EW}^2/M_{\rm GUT})\sim O(10^{-3})$ eV. 
The current experimental data for neutrino masses is
$\Delta m_{21}^2=(7.50\pm 0.20)\times 10^{-5}$ eV$^2$ and
$\Delta m_{32}^2=0.00232^{+0.00012}_{-0.00008}$ eV$^2$
\cite{Beringer:1900zz}, so it seems better that the mediated particles 
have smaller mass $O(10^{14})-O(10^{15})$ GeV than GUT-scale mass 
$M_{\rm GUT}\sim O(10^{16})$ GeV. 

For $SU(5)$ GUT models, Type-I, Type-II, and Type-III seesaw mechanisms
can be also achieved  by using $SU(5)$ singlet fields $\hat{N}_{\bf 1}$
$SU(5)$ ${\bf 15}$-plet and ${\bf 15}^*$-plet fields 
$\hat{T}_{\bf 15}$ and $\hat{T}_{{\bf 15}^*}^c$,
and $SU(5)$ ${\bf 24}$-plet fields $\hat{A}_{\bf 24}$, respectively.
Note that since $\hat{A}_{\bf 24}$ contains $\hat{A}_\ell$ and
$\hat{W}_\ell$, this field includes not only Type-I seesaw but
also Type-III seesaw mechanisms.

We move on to our $SU(5)\times SU(1,1)$ model. As we have already seen
before, the Majorana mass terms are not allowed by the $SU(1,1)$
symmetry. One may think that the Type-I and Type-III seesaw mechanisms
are prohibited, but as we discussed for the effective $\mu$ term of the
up- and down-type higgs doublets, once the horizontal symmetry is broken,
there is no reason to prohibit the Majorana mass terms.

We discuss two situations Majorana-type Type-I and Type-III seesaw
mechanisms and Dirac-type Type-I, Type-II, and Type-III seesaw
mechanisms for the massive mediated superfields realized by the
spontaneous generations of generations discussed in Sec.~\ref{Sec:SGG}.

We start by considering the Majorana-type Type-I and Type-III seesaw
mechanisms. The masses of the mediated fields come from the Dirac mass
term of the fields and their conjugate fields, and the masses are
different from the Majorana masses $\mu_X$, where $\mu_X$ stands for 
the Majorana mass of $\hat{N}_{\bf 1}$ or $\hat{A}_{\bf 24}$. Our basic
assumption is that $\mu_X$ is much smaller than the Dirac mass of the
the mediated fields. When we integrate out the mediated fields, we obtain 
the effective neutrino masses $O(\mu_X v_{\rm EW}^2/M_{\rm GUT}^2)$.
If we assume $\mu_N\sim O(\sqrt{m_{\rm SUSY}M_{\rm GUT}})$, 
$O(\mu_N v_{\rm EW}^2/M_{\rm GUT}^2)\sim O(10^{-10})$ eV.
This is too tiny. Therefore, the Majorana-type seesaw mechanisms cannot
explain the observed neutrino masses.

Next we discuss Dirac-type Type-I, Type-II, and Type-III seesaw
mechanisms. The superpotential terms are given by
\begin{align}
W_{\rm D-I}=&
y_{I}\hat{G}_{{\bf 5}^*}\hat{H}_{u{\bf 5}}\hat{N}_{\bf 1}
+y_{I}^c\hat{G}_{{\bf 5}^*}'\hat{H}_{u{\bf 5}}\hat{N}_{\bf 1}^c,\\
W_{\rm D-II}=&
y_{II}\hat{G}_{{\bf 5}^*}\hat{G}_{{\bf 5}^*}'\hat{T}_{\bf 15}
+y_{II}^c\hat{H}_{u{\bf 5}}\hat{H}_{u{\bf 5}}\hat{T}_{{\bf 15}^*}^c,\\
W_{\rm D-III}=&
y_{III}\hat{G}_{{\bf 5}^*}\hat{H}_{u{\bf 5}}\hat{A}_{\bf 24}
+y_{III}^c\hat{G}_{{\bf 5}^*}'\hat{H}_{u{\bf 5}}\hat{A}_{\bf 24}^c,
\end{align}
where $\gamma=\beta+\xi+\Delta_\xi$, $\beta'=\gamma+\xi+\Delta_\xi'$,
$\tau=\beta+\beta'+\Delta_\tau$, $\tau=2\gamma+\Delta_\tau'$,
$\gamma=\beta+\zeta+\Delta_\zeta$, and 
$\beta'=\gamma+\zeta+\Delta_\zeta'$. $\Delta$s are non-negative integer.
To realize the seesaw mechanisms, we have to choose the $SU(1,1)$ weight
assignment satisfying the following condition:
\begin{align}
2\gamma=\beta+\beta'+\Delta_{x}-\Delta_{x}',
\end{align}
where $x$ stands for $\xi$, $\tau$, or $\zeta$.
This leads to a constraint 
\begin{align}
n=4m+\Delta_x-\Delta_x'-2q_\alpha+q_\beta-\frac{1}{2}
\end{align}
for the $SU(1,1)$ weight assignment in Eq.~(\ref{Assignment:SU(1,1)}),
and also leads to a constraint 
\begin{align}
n=4m+\Delta_x-\Delta_x'-10q_\alpha-q_\beta-\frac{5}{2}
\end{align}
for the $SU(1,1)$ weight assignment in Eq.~(\ref{Assignment:SU(1,1)-2}).
After decoupling the heavy matter, we obtain the effective superpotential
\begin{align}
W_{N}=\sum_{n,m=0}^2\kappa_{n,m}^X
\hat{\ell}_n\hat{\ell}_m\hat{h}_u\hat{h}_u,
\end{align}
where $\kappa_{n,m}^X$ are coupling constants, and $X$ stands for I,
II, and III,
\begin{align}
\kappa_{n,m}^{I}:=&
\sum_{i,j,k=0}^{\infty}
-\frac{y_Iy_I^c}{\tilde{M}_j^I}
C_{i,j}^{\beta,\xi,\Delta_\xi}C_{k,j}^{\gamma,\xi,\Delta_\xi'}
U_{n,i}^{\ell}U_{m,k+j-\Delta_\xi'}^{\ell'}
U_{i+j-\Delta_\xi}^{h_u}U_k^{h_u},
\label{Neutrino-mass-I}\\
\kappa_{n,m}^{II}:=&
\sum_{i,j,k=0}^{\infty}
-\frac{y_{II}y_{II}^c}{\tilde{M}_j^{II}}
C_{i,i+j-\Delta_\tau}^{\beta,\beta',\Delta_\tau}
C_{k,j-k+\Delta_\tau'}^{\gamma,\gamma,\Delta_\tau'}
U_{n,i}^{\ell}U_{m,j-i+\Delta_\tau}^{\ell'}
U_{k}^{h_u}U_{j-k+\Delta_\tau'}^{h_u},\\
\kappa_{n,m}^{III}:=&
\sum_{i,j,k=0}^{\infty}
-\frac{y_{III}y_{III}^c}{\tilde{M}_j^{III}}
C_{i,j}^{\beta,\zeta,\Delta_\zeta}C_{k,j}^{\gamma,\zeta,\Delta_\zeta'}
U_{n,i}^{\ell}U_{m,k+j-\Delta_\zeta'}^{\ell'}
U_{i+j-\Delta_\zeta}^{h_u}U_k^{h_u},
\label{Neutrino-mass-III}
\end{align}
where $\tilde{M}_j^{X}$ is the mass of the $j$th component of the
Type-I, II, III. 

Here we comment on neutrino masses of Type-I, II, and III in
Eqs.~(\ref{Neutrino-mass-I})--(\ref{Neutrino-mass-III}), qualitatively.
The VEV $\langle h_u\rangle=O(v_{\rm EW})$ of the chiral up-type higgs
$\hat{h}_u$ generates the effective neutrino mass matrix 
$m_\nu^{mn}\hat{\ell}_m\hat{\ell}_n~(m,n=0,1,2)$ with
$m_\nu^{mn}=O(v_{\rm EW}^2/M_{\rm GUT})$.  
To realize the observed neutrino masses, the mass of a mediated particle
should have a smaller value compared with $M_{\rm GUT}=O(10^{16})$
GeV. Unfortunately, to obtain the smaller mass, we need some fine-tuning
between the original mass of the mediated particle $O(M_{\rm GUT})$ and
the nonvanishing VEV of the structure fields $O(M_{\rm GUT})$. 
Here we assume that the components surrounded by the circles 
in Figs.~\ref{fig:SGG-new-p=3/2-T1}, \ref{fig:SGG-new-p=3/2-T2} and
\ref{fig:SGG-new-p=3/2-T3} are the main elements.
For Type-I, the neutrino masses appear only in two elements at the
leading order. For small sub-leading contribution derived from
$\langle\phi_0\rangle$ and $\langle\phi_{+1}'\rangle$, the predicted
masses seem to be incompatible with the observed masses. 
In principle, the sub-leading contribution derived from 
$\langle\phi_0\rangle$ and $\langle\phi_{+1}'\rangle$ can be large, so
it may reproduce the observed masses. 
For Type-II and III, the neutrino masses vanish for leading order,
i.e., a limit $\langle\phi_{0}\rangle$ and $\langle\phi_{+1}'\rangle$
going to zero. Thus, the sub-leading component mainly contribute to
the neutrino masses. In this case, since the overall coupling becomes
small, we need finer tuning to realize the observed neutrino masses.
When the other contribution is small, the neutrino mass matrix seems
to be normal hierarchy. When the other contribution is large, it
depends on parameters.

\subsection{Proton decay}

Before we discuss proton decay in our model, we quickly
review the proton decay discussion in the minimal $SU(5)$ SUSY GUT
\cite{Sakai:1981pk,Weinberg:1981wj,Nath:1985ub,Goto:1998qg}.
First, the superpotential of the Yukawa couplings in models with the
minimal $SU(5)$ matter content contains the following baryon and/or
lepton number violation terms 
\begin{align}
W=&y_{\bf 10}\hat{F}_{\bf 10}\hat{F}_{\bf 10}\hat{H}_{u{\bf 5}}
\ni
y_{\bf 10}\hat{Q}\hat{Q}\hat{T}_{u}
+y_{\bf 10}\hat{E}^c\hat{U}^c\hat{T}_{u},\\
W=&y_{\bf 10}\hat{F}_{\bf 10}\hat{G}_{{\bf 5}^*}\hat{H}_{d{\bf 5}^*}
\ni
y_{\bf 5}\hat{Q}\hat{L}\hat{T}_{d}
+y_{\bf 5}\hat{U}^c\hat{D}^c\hat{T}_{d}.
\end{align}
After the doublet part of the original $SU(5)$ $\mu$-term 
$\mu_{\bf 5}\hat{H}_{u{\bf 5}}\hat{H}_{d{\bf 5}}$ is canceled by using
the ``$\mu$''-term induced from the VEVs of the coupling between the
$SU(5)$ adjoint and up- and down-type higgses
$\langle\hat{\Phi}_{\rm 24}\rangle\hat{H}_{u{\bf 5}}\hat{H}_{d{\bf 5}^*}$,
we can obtain the effective $\mu$ parameter of the doublet higgses
$\mu\sim O(m_{\rm SUSY})$ and of the colored higgses 
$M_C\sim O(M_{\rm GUT})$.
After the colored higgses decouple, they lead to two 
superpotential terms that include dimension-5 operators breaking baryon
and/or lepton number 
\begin{align}
W_{5}=-\frac{1}{M_C}\sum_{m,n,p,q=0}^2
\left(\frac{1}{2}
 C_{5L}^{mnpq}\hat{q}_m  \hat{q}_n  \hat{q}_p  \hat{\ell}_q
+C_{5R}^{mnpq}\hat{e}_m^c\hat{u}_n^c\hat{u}_p^c\hat{d}_q^c
\right),
\label{Eq:B-L-term}
\end{align}
where $C_{5X}^{mnpq}$ $(X=L,R)$ are dimensionless coupling constants
that depend on the Yukawa coupling matrices of quarks and leptons.
According to the analysis discussed in Ref.~\cite{Goto:1998qg}, 
we use the recent super-Kamiokande result for the lifetime 
$\tau(p\to K^+\bar{\nu})>3.3\times 10^{33}$ years at 90\% C.L.
\cite{Miura:2010zz}.
Assuming that soft SUSY breaking parameters at the Planck scale
are described by the universal scalar mass, universal gaugino mass, and
universal coefficient of the trilinear scalar coupling, so-called
$A$-term and the sfermion mass $m_{\tilde{f}}$ is less than $1$ TeV, 
the colored higgs mass $M_C$ must be larger than  $10^{17}$ GeV for
$\tan\beta$ $(2<\tan\beta<5)$; $10^{18}$ GeV for $\tan\beta=10$;
$10^{19}$ GeV for $\tan\beta=30$; and $10^{20}$ GeV for $\tan\beta=50$.
(Recently, it was discussed in 
Ref.~\cite{McKeen:2013dma,Liu:2013ula,Hisano:2013kl} that when the
sfermion mass is much greater than $1$ TeV, the colored higgs mass $M_C$
can be $10^{16}$ GeV regardless of $\tan\beta$.) 

We move on to discuss proton decay in our model. 
The chiral matter content is realized via the spontaneous
generation of generations discussed in Sec.~\ref{Sec:SGG}. 
As discussed in Sec.~\ref{Sec:SGG:Higgs}, once the up-
and down-type doublet higgses appear and the up- and down-type colored
higgses disappear at a vacuum, the up- and down-type colored higgses
have their Dirac masses.
To generate the baryon and/or lepton number violation terms 
in Eq.~(\ref{Eq:B-L-term}), they must include the $\mu$-term
between the colored higgses.
We discuss two assignments in Eqs.~(\ref{Assignment:SU(1,1)}) and
(\ref{Assignment:SU(1,1)-2}). 
The effective superpotential is 
\begin{align}
W=\sum_{m,n,p,q=0}^2
\lambda_{m,n,p,q}
\hat{q}_m\hat{q}_n\hat{q}_p\hat{\ell}_q
+\lambda_{m,n,p,q}'\hat{e}_m^c\hat{u}_n^c\hat{u}_p^c\hat{d}_q^c,
\label{W:proton-decay-SU(1,1)}
\end{align}
where $\lambda_{m,n,p.q}$ and $\lambda_{m,n,p,q}'$ are determined by the
colored higgs 
masses, the $\mu$-parameter of the colored higgses, the overall Yukawa
couplings, and the mixing coefficients of quarks and leptons.
For the assignment in Eq.~(\ref{Assignment:SU(1,1)}), the Yukawa
couplings in Eqs.~(\ref{W:Yukawa-10-H}) and (\ref{W:Yukawa-5-H}) lead to 
\begin{align}
\lambda_{m,n,p,q}
=&\sum_{i,j,k,\ell=0}^{\infty}
\bigg[
y_{\bf 10}y_{\bf 5}
\frac{\mu_{i+j,k+\ell}}
{M_{t_u i+j}M_{t_d k+\ell}}
C_{i,j}^{\alpha,\alpha',0}C_{k,\ell}^{\alpha',\beta,0}
U_{m,i}^qU_{n,j}^{q'}U_{p,k}^{q'}U_{q,\ell}^{\ell}\nonumber\\
&+y_{\bf 10}y_{\bf 5}'
\frac{\mu_{i+j,k+\ell-q_\beta+q_\alpha}}
{M_{t_u i+j}M_{t_d k+\ell-q_\beta+q_\alpha}}
C_{i,j}^{\alpha,\alpha',0}C_{k,\ell}^{\alpha,\beta',q_\beta-q_\alpha}
U_{m,i}^qU_{n,j}^{q'}U_{p,k}^{q}U_{q,\ell}^{\ell'}\bigg],
\end{align}
\begin{align}
\lambda_{m,n,p,q}'
=&\sum_{i,j,k,\ell=0}^{\infty}
\bigg[
y_{\bf 10}y_{\bf 5}
\frac{\mu_{i+j,k+\ell}}
{M_{t_u i+j}M_{t_d k+\ell}}
C_{i,j}^{\alpha,\alpha',0}C_{k,\ell}^{\alpha',\beta,0}
U_{m,i}^eU_{n,j}^{u'}U_{p,k}^{u'}U_{q,\ell}^{d}\nonumber\\
&+y_{\bf 10}y_{\bf 5}'
\frac{\mu_{i+j,k+\ell-q_\beta+q_\alpha}}
{M_{t_u i+j}M_{t_d k+\ell-q_\beta+q_\alpha}}
C_{i,j}^{\alpha,\alpha',0}C_{k,\ell}^{\alpha,\beta',q_\beta-q_\alpha}
U_{m,i}^{e}U_{n,j}^{u'}U_{p,k}^{u}U_{q,\ell}^{d'}\bigg],
\end{align}
where 
\begin{align}
\mu_{i,j}:=
\mu\left(\prod_{r=0}^{i-1}\frac{M_{t_u r}'}{M_{t_u r}}\right)
\left(\prod_{s=0}^{j-1}\frac{M_{t_d s}'}{M_{t_d s}}\right)
\end{align}
and
\begin{align}
M_{t_u i}:=&M_{h_u}(-1)^i+x_{h_u}\langle\phi_0\rangle
D_{i,i}^{\gamma,\gamma,S},\ \ \
M_{t_u i}':=Y_{t_u}z_{h_u}\langle\phi_{+1}'\rangle
D_{i+1,i}^{\gamma,\gamma,S'},\nonumber\\
M_{t_d i}:=&M_{h_d}(-1)^i+x_{h_d}\langle\phi_0\rangle
D_{i,i}^{\delta,\delta,S},\ \ \
M_{t_d i}':=Y_{t_d}z_{h_d}\langle\phi_{+1}'\rangle
D_{i+1,i}^{\delta,\delta,S'}.
\end{align}
For the assignment in Eq.~(\ref{Assignment:SU(1,1)-2}), the Yukawa
couplings in Eqs.~(\ref{W:Yukawa-10-I}) and (\ref{W:Yukawa-5-I}) lead to
\begin{align}
\lambda_{m,n,p,q}
=&\sum_{i,j,k,\ell=0}^{\infty}
\bigg[
y_{\bf 10}y_{\bf 5}
\frac{\mu_{i+j-2q_\alpha,k+\ell-q_\alpha-q_\beta}}
{M_{t_u i+j-2q_\alpha}M_{t_d k+\ell-q_\alpha-q_\beta}}
C_{i,j}^{\alpha,\alpha,2q_\alpha}
C_{k,\ell}^{\alpha,\beta,q_\alpha+q_\beta}
U_{m,i}^qU_{n,j}^{q}U_{p,k}^{q}U_{q,\ell}^{\ell}\nonumber\\
&+y_{\bf 10}y_{\bf 5}'
\frac{\mu_{i+j-2q_\alpha,k+\ell}}
{M_{t_u i+j-2q_\alpha}M_{t_d k+\ell}}
C_{i,j}^{\alpha,\alpha,2q_\alpha}C_{k,\ell}^{\alpha',\beta',0}
U_{m,i}^qU_{n,j}^{q}U_{p,k}^{q'}U_{q,\ell}^{\ell'}\nonumber\\
&+y_{\bf 10}'y_{\bf 5}
\frac{\mu_{i+j,k+\ell-q_\alpha-q_\beta}}
{M_{t_u i+j}M_{t_d k+\ell-q_\alpha-q_\beta}}
C_{i,j}^{\alpha',\alpha',0}C_{k,\ell}^{\alpha,\beta,q_\alpha+q_\beta}
U_{m,i}^{q'}U_{n,j}^{q'}U_{p,k}^{q}U_{q,\ell}^{\ell}\nonumber\\
&+y_{\bf 10}'y_{\bf 5}'
\frac{\mu_{i+j,k+\ell}}
{M_{t_u i+j}M_{t_d k+\ell}}
C_{i,j}^{\alpha',\alpha',0}C_{k,\ell}^{\alpha',\beta',0}
U_{m,i}^{q'}U_{n,j}^{q'}U_{p,k}^{q'}U_{q,\ell}^{\ell'}
\bigg],
\end{align}
\begin{align}
\lambda_{m,n,p,q}'
=&\sum_{i,j,k,\ell=0}^{\infty}
\bigg[
y_{\bf 10}y_{\bf 5}
\frac{\mu_{i+j-2q_\alpha,k+\ell-q_\alpha-q_\beta}}
{M_{t_u i+j-2q_\alpha}M_{t_d k+\ell-q_\alpha-q_\beta}}
C_{i,j}^{\alpha,\alpha,2q_\alpha}
C_{k,\ell}^{\alpha,\beta,q_\alpha+q_\beta}
U_{m,i}^eU_{n,j}^{u}U_{p,k}^{u}U_{q,\ell}^{d}\nonumber\\
&+y_{\bf 10}y_{\bf 5}'
\frac{\mu_{i+j-2q_\alpha,k+\ell}}
{M_{t_u i+j-2q_\alpha}M_{t_d k+\ell}}
C_{i,j}^{\alpha,\alpha,2q_\alpha}C_{k,\ell}^{\alpha',\beta',0}
U_{m,i}^{e}U_{n,j}^{u}U_{p,k}^{u'}U_{q,\ell}^{d'}\nonumber\\
&+y_{\bf 10}'y_{\bf 5}
\frac{\mu_{i+j,k+\ell-q_\alpha-q_\beta}}
{M_{t_u i+j}M_{t_d k+\ell-q_\alpha-q_\beta}}
C_{i,j}^{\alpha',\alpha',0}C_{k,\ell}^{\alpha,\beta,q_\alpha+q_\beta}
U_{m,i}^{e'}U_{n,j}^{u'}U_{p,k}^{u}U_{q,\ell}^{d}\nonumber\\
&+y_{\bf 10}'y_{\bf 5}'
\frac{\mu_{i+j,k+\ell}}
{M_{t_u i+j}M_{t_d k+\ell}}
C_{i,j}^{\alpha',\alpha',0}C_{k,\ell}^{\alpha',\beta',0}
U_{m,i}^{e'}U_{n,j}^{u'}U_{p,k}^{u'}U_{q,\ell}^{e'}
\bigg].
\end{align}
Note that as we discussed in Sec.~\ref{Sec:mu-term}, the original $\mu$
term between up- and down-type higgses is prohibited by the horizontal
symmetry. In the model, the nonvanishing VEVs of the singlets generate
the $\mu$-term of the $0$th component of $\hat{h}_{u{\bf 5}}$ and
$\hat{h}_{d{\bf 5}^*}$.

We need to consider the experimental bound for proton decay in the
model. In the calculation in Ref.~\cite{Goto:1998qg} it is assumed that
the Yukawa coupling matrices of the colored higgses are the same as the
matrices of the down-type higgses. In the current model, the Yukawa
coupling matrices of the doublet higgses are different from the colored
higgses, so we cannot use directly the constraint for the mass of the
colored higgses discussed in Ref.~\cite{Goto:1998qg}.  
When we assume that the $C_{5X}^{mnpq}$ in Eq.~(\ref{Eq:B-L-term}) is 
almost the same as the coupling constant in
Eq.~(\ref{W:proton-decay-SU(1,1)}) normalized by using the effective  
mass $M_{t_u 0}M_{t_d 0}/\mu_{0,0}$, and we compare
Eq.~(\ref{Eq:B-L-term}) with the above coupling constants, 
the value of the effective $M_C$ is $M_{t_u 0}M_{t_d 0}/\mu_{0,0}$. 
When we assume $M_{t_u 0}\sim M_{t_d 0}\sim M_{\rm GUT}\sim O(10^{16})$ GeV
and  $\mu_{0,0}\sim m_{\rm SUSY}\sim O(10^3)$, we obtain 
$M_C\sim O(10^{29})$ GeV. 
This value is far from the current colored higgs mass bound $O(10^{17})$
GeV to $O(10^{20})$ GeV. Thus, the proton decay effect caused by the
colored higgs is negligible once the colored higgs are massive via the
spontaneous generation of generations. The dominant contribution for
proton decay modes comes from the $X$ and $Y$ gauge boson exchanges. 
The dominant proton decay mode $p\to\pi^0e^+$ via the $X$ and $Y$ gauge
bosons must be found first. In other words, if one of the current or
planned near future proton decay experiments finds another proton decay
mode, e.g., $p\to K^+\bar{\nu}$ before $p\to\pi^0e^+$ are found, this
model will be excluded.  

Finally, we verify the contribution from additional matter fields
$\hat{C}_h$ and $\hat{Q}_h$ in the ${\bf 15}$-plet $\hat{T}_{\bf 15}$,
and $\hat{X}_\ell$ and $\hat{Y}_\ell$ in the ${\bf 24}$-plet
$\hat{A}_{\bf 24}$, and their conjugate fields. These terms cannot
generate the superpotential quartic terms in
Eq.~(\ref{W:Proton-decay}), so the lowest contribution can only come
from at least superpotential quintic terms. 
This means that since the nonvanishing VEVs of the non-SM singlets are
those of up- and down-type higgses, the contribution of them for proton
decay are suppressed by at least 
$m_{\rm SUSY}/M_{\rm GUT}\sim O(10^{-13})$ compared to the
superpotential quartic terms in
Eq.~(\ref{W:Proton-decay}). Thus, they are completely negligible at 
least for the current experimental bound.

\section{Summary and discussion}
\label{Sec:Summary}

We discussed the $SU(5)$ SUSY GUT model with the $SU(1,1)$ horizontal
symmetry that includes the matter fields in Table \ref{tab:matter}
and the structure fields in Table \ref{tab:structure}.
We showed that the mechanism of the spontaneous generation of
generations produces the matter content of the MSSM and the almost
decoupled $G_{\rm SM}$ singlets through the nonvanishing VEVs of the
structure fields given in Eq.~(\ref{Structure-VEVs}).
For quarks and leptons, the nonvanishing VEV
$\langle\psi_{-3/2}\rangle$ of the structure field 
$\hat{\Psi}_{{\bf 1}/{\bf 24}}$ with the $SU(1,1)$ half-integer spin 
$S''$ plays the important role for producing the three chiral
generations of quarks and leptons. 
The nonvanishing VEV $\langle\phi_{+1}'\rangle$ of the structure field
$\hat{\Phi}_{\bf 24}'$ with the $SU(1,1)$ integer $S'$ leads to the
difference between the mixing coefficients of quarks and leptons
because the structure field $\hat{\Phi}_{\bf 24}'$ belongs to the
nontrivial representation of $SU(5)$. 
Thus, the mixing coefficients of the down-type quarks are different
from those of the charged leptons. This avoids the unacceptable
prediction in the minimal $SU(5)$ GUT model for the down-type quark's
and the charged lepton's Yukawa coupling constants.
For higgses, the nonvanishing VEV $\langle\psi_{-3/2}\rangle$
does not affect anything because the structure field 
$\hat{\Psi}_{{\bf 1}/{\bf 24}}$ does not couple to the higgs superfields.
Due to this fact, the nonvanishing VEVs $\langle\phi_{0}\rangle$ and 
$\langle\phi_{+1}'\rangle$ of the structure fields
$\hat{\Phi}_{{\bf 1}}$ and $\hat{\Phi}_{{\bf 24}}'$ 
with the $SU(1,1)$ integers $S$ and $S'$ determine whether the higgses
appear or not. The VEVs can produce only one generation of the up- and
down-type doublet higgses at low energy. 
We found that the model naturally realizes the doublet-triplet mass
splitting between the doublet and colored higgses pointed out 
in Ref.~\cite{Inoue:2000ia,Yamatsu:2007}.

We also found that some special $SU(1,1)$ assignments allow only 
the $\cancel{B}$ and/or $\cancel{L}$ superpotential quartic term 
$\hat{G}_{{\bf 5}^*}\hat{G}_{{\bf 5}^*}'\hat{H}_{u{\bf 5}}\hat{H}_{u{\bf
5}}$, 
which contains the $\hat{L}\hat{L}'\hat{H}_u\hat{H}_u$, up to
superpotential quartic order.
The assignments retain $R$-parity even after the $SU(1,1)$ symmetry is
broken. Thus, we can identify the $SU(1,1)$ assignments as the origin of
the $R$-parity. 

We found that this model can generate the neutrino masses via not only
the Type-II seesaw mechanism but also the Type-I and Type-III seesaw
mechanisms. We also found that the neutrino masses are dependent on
the mixing coefficients of the leptons and up-type higgses,
the $SU(1,1)$ CGCs, the masses of the mediated fields, and their overall
Yukawa coupling constants.

We verified that the proton decay induced via the superpotential quartic
terms generated by decoupling the colored higgses is highly
suppressed compared to that of usual GUT models. The suppression factor
is roughly $O(m_{\rm SUSY}/M_{\rm GUT})\sim O(10^{-13})$. Thus, the 
dominant contribution to proton decay comes from the $X$ and $Y$ gauge
bosons. Thus, the dominant proton decay mode $p\to\pi^0e^+$ via the $X$
and $Y$ gauge bosons must be found first. In other words, if
another proton decay mode, e.g., $p\to K^+\bar{\nu}$, is discovered
before $p\to\pi^0e^+$ is found, this model will be excluded.

We mention the gauge anomalies of $G_{SM}$ at low energies.
The spontaneous generation of generations allows apparent anomalous
chiral matter content at low energies because the apparent anomalies
should be canceled out by the Wess-Zumino-Witten term
\cite{Wess:1971yu,Witten:1983tw,Yamatsu:2008}. 
For example, the up-type colored higgs could appear at low energy while
the down-type colored higgs disappears at low energy. In this case, the
matter content at low energy is anomalous. Of course, since in this
situation there is a massless colored higgsino, this is unacceptable.
The apparent anomaly cancellation exhibited by the observed low energy
fields therefore appears coincidental in some sense if the spontaneous
generation of generations is realized in nature.

We also mention ``charge'' quantization of weights of $SU(1,1)$ in this
model. The $SU(1,1)$ spins of structure fields are obviously quantized
because of finite-dimensional representations of $SU(1,1)$, while
the lowest(highest) $SU(1,1)$ weight of matter fields are arbitrary 
and there is no reason to quantize their ``charges.''
Of course, we need the ``charge'' quantization for matter fields, e.g.,
to  realize three chiral generations of quarks and leptons at 
low-energy and the existence of Yukawa couplings.
The charge quantization may be realized naturally in part if we embed
$SU(1,1)$ into a higher rank noncompact group, e.g., $SU(2,1)$.
The unitary representations of $SU(2,1)$ live on a two dimensional plain.
The generators of $SU(2,1)$ can be written by three dependent
subgroups two $SU(1,1)$ and one $SU(2)$, just like $SU(3)$ that
can be written by three dependent subgroups $SU(2)$.
Still, the charges of the lowest state of the $SU(2,1)$
representations are arbitrary, but since the unitary representations
of $SU(2,1)$ contain the representations of $SU(1,1)$ with different
weights by integer times a certain fraction, the difference between the
charges of the lowest state of the $SU(1,1)$ can be quantized. 
At present, since there are no works to discuss models with a higher
rank noncompact group horizontal symmetry, it has not been discovered 
when and how the spontaneous generation of generations works.

We have not yet solved the vacuum structure in a model that includes
at least three structure fields with two $SU(1,1)$ integer spins $S$,
$S'$ and one $SU(1,1)$ half-integer spin $S''$. To produce three chiral
generations of quarks and leptons and one generation of higgses, the
$SU(1,1)$ spins must satisfy the relation $S''>S=S'\geq 1$.
Thus, the minimal choice is $S=S'=1$, $S''=3/2$. We must discuss the
model to justify the assumption of this article.

We comment on nonrenormalizable terms when they are generated by Planck 
scale physics. For matter fields, as we discussed in
Sec.~\ref{Sec:BLBreaking}, since special weight assignments of 
$SU(1,1)$ allow only the superpotential term in
Eq.~(\ref{W:Neutrino-mass}) up to quartic order, 
the effect does not seem to affect anything at low energy.
We have problems if higher order terms between structure and
matter fields are generated by Planck scale physics.
For example, let us consider a model that includes a matter field 
$\hat{F}$, its conjugate field $\hat{F}^c$ and a structure field
$\hat{\Phi}$ with an $SU(1,1)$ integer spin $S$, where 
we assume that the $g$th component of the structure field $\hat{\Phi}$
has a nonvanishing VEV. 
The relevant superpotential terms for the spontaneous generation of
generations are 
\begin{align}
W=M\hat{F}\hat{F}^c+
\sum_{m=1}^{\ell}\frac{C_m}{\Lambda^{m}}\hat{F}\hat{F}^c\hat{\Phi}^m,
\end{align}
where $M$ is a mass parameter, $C_m$s are dimensionless coupling
constants, $\Lambda$ is a Planck scale mass parameter,
and $\ell$ is an integer number. $\ell g$th generations of the massless
modes $\hat{f}_n$ $(n=0,1,\cdots,\ell g-1)$ appear because the largest
spin state built by $\hat{\Phi}^m$ has the spin $mS$ and this coupling is
the dominant contribution to produce the chiral particles regardless of
coupling constants $C_m$s. 
From the viewpoint of effective theory, there is no reason that $\ell$
is finite. For $\ell\to\infty$, the number of chiral generations
is zero for $g=0$ and $\infty$ for $g\not=0$.
At present, we must assume that unknown fundamental theory 
only allows renormalizable terms of the structure and matter field
sector to justify our discussion.

\section*{Acknowledgments}

The author would like to thank Jonathan P. Hall, Kenzo Inoue, 
Hirofumi Kubo, and Seodong Shin for helpful comments.
This work was supported in part by the Department of Energy under grant
DE-FG02-91ER40661 and by the Indiana University Center for Spacetime
Symmetries.

\bibliographystyle{utphys} 
\bibliography{reference}

\end{document}